

\input harvmac.tex
\input epsf

\def\figin{\epsfcheck\figin}\def\figins{\epsfcheck\figins}
\def\epsfcheck{\ifx\epsfbox\UnDeFiNeD
\message{(NO epsf.tex, FIGURES WILL BE IGNORED)}
\gdef\figin##1{\vskip2in}\gdef\figins##1{\hskip.5in}
\else\message{(FIGURES WILL BE INCLUDED)}%
\gdef\figin##1{##1}\gdef\figins##1{##1}\fi}
\def\DefWarn#1{}
\def\figinsert{\goodbreak\midinsert}
\def\ifig#1#2#3{\DefWarn#1\xdef#1{fig.~\the\figno}
\writedef{#1\leftbracket fig.\noexpand~\the\figno}%
\figinsert\figin{\centerline{#3}}\medskip\centerline{\vbox{\baselineskip12pt
\advance\hsize by -1truein\noindent\footnotefont{\bf Fig.~\the\figno:} #2}}
\bigskip\endinsert\global\advance\figno by1}

\def\Title#1#2{\rightline{#1}
\ifx\answ\bigans\nopagenumbers\pageno0\vskip1in%
\baselineskip 15pt plus 1pt minus 1pt
\else
\pageno1\vskip.4in\fi \centerline{\titlefont #2}\vskip .4in}

\ifx\answ\bigans\def\tcbreak#1{}\else\def\tcbreak#1{\cr&{#1}}\fi

\def\ee{{\rm e}}
\font\cmss=cmss10 \font\cmsss=cmss10 at 7pt
\def\IZ{\relax\ifmmode\mathchoice
{\hbox{\cmss Z\kern-.4em Z}}{\hbox{\cmss Z\kern-.4em Z}}
{\lower.9pt\hbox{\cmsss Z\kern-.4em Z}}
{\lower1.2pt\hbox{\cmsss Z\kern-.4em Z}}\else{\cmss Z\kern-.4em Z}\fi}

\lref\REFwitmas{E. Witten, Nucl. Phys. {\bf B149} (1979) 285.}
\lref\REFmigrev{A. A. Migdal, Physics Reports {\bf 102} (1983) 199.}
\lref\REFyaffe{L. Yaffe, Rev. Mod. Phys. {\bf 54} (1982) 407.}
\lref\REFmami{Yu. Makeenko and A. A. Migdal, Nucl. Phys. {\bf B188}
 (1981) 269.}
\lref\REFkazkos{V. A. Kazakov and I. K. Kostov, Nucl. Phys. {\bf B176}
(1980) 199.}
\lref\REFdaulkaz{D. V. Boulatov, Mod. Phys. Lett. {\bf A9} (1994) 365\semi
J.-M. Daul and V. A. Kazakov, {\sl Wilson Loop
for Large $N$ Yang--Mills Theory on a Two-Dimensional Sphere},
preprint LPTENS-9337, hep-th/9310165.}
\lref\REFdou{M. R. Douglas, {\sl Conformal Field Theory Techniques
in Large $N$ Yang--Mills Theory}, hep-th/9311130.}
\lref\REFmigdal{A. A. Migdal, Sov. Phys. JETP {\bf 42} (1975) 413 \
(Zh. Exp. Teor. Fiz. {\bf 69} (1975) 810).}
\lref\REFrusakov{B. Rusakov, Mod. Phys. Lett. {\bf A5} (1990) 693.}
\lref\REFdouglas{M. R. Douglas and V. A. Kazakov, Phys. Lett. {\bf 319B}
(1993) 219, hep-th/9305047.}
\lref\REFtaylor{D. Gross, Nucl. Phys. {\bf B400} (1993) 161\semi
D. Gross and W. Taylor, Nucl. Phys. {\bf B400} (1993) 181\semi
Nucl. Phys. {\bf B403} (1993) 395.}
\lref\REFminahan{J. Minahan and A. Polychronakos, {\sl Classical Solutions
for Two-Dimensional QCD on the Sphere}, preprint CERN-TH-7016/93,
UVA-HET-93-08, hep-th/9309119.}
\lref\REFmincone{J. Minahan and A. Polychronakos, Phys. Lett. {\bf 312B}
(1993) 155.}
\lref\REFDADDA{A. D'Adda, M. Caselle, L. Magnea  and S. Panzeri,
{\sl Two dimensional QCD on the sphere and on the Cylinder}, hep-th/9309107.}
\lref\REFgrwit{D. Gross and E. Witten, Phys. Rev. {\bf D21} (1980) 446.}
\lref\REFinst{D. Gross and A. Matytsin, {\sl Instanton Induced Large $N$
Phase Transitions in Two and Four Dimensional QCD}, preprint PUPT-1459,
hep-th/9404004.}
\lref\REFjesa{A. Jevicki and B. Sakita, Nucl. Phys. {\bf B165} (1980) 511\semi
S. R. Das and A. Jevicki, Mod. Phys. Lett. {\bf A5} (1990) 1639.}
\lref\REFizube{C. Itzykson and J.-B. Zuber, Jour. Math. Phys. {\bf 21}
(1980) 411.}
\lref\REFtho{G. 'tHooft, Nucl. Phys. {\bf B72} (1974) 461\semi
Nucl. Phys. {\bf B75} (1974) 461.}
\lref\REFmat{A. Matytsin, Nucl. Phys. {\bf B411} (1994) 805.}
\lref\REFrossi{P. Rossi, Ann. Phys. {\bf 132} (1981) 463.}
\lref\REFdgtr{M. R. Douglas, {\sl Large $N$ Gauge Theory -- Expansions and
Transitions}, preprint RU-94-72,
hep-th/9409098.}
\lref\REFdol{B. Durnhuus and P. Olesen, Nucl. Phys. {\bf B184} (1981) 161.}
\lref\REFwitte{E. Witten, Commun. Math. Phys. {\bf 141} (1991) 153\semi
J. Geom. Phys. {\bf 9} (1992) 303.}

\Title{\vbox{\baselineskip12pt\hbox{PUPT-1503}
\hbox{hep-th/9410054}}}
{\vbox{\centerline{Some Properties of Large $N$}\vskip0.15in
\centerline{Two Dimensional Yang--Mills Theory}}}
\centerline{David J. Gross\footnote{$^\dagger$}
{This work was supported in part by the National Science Foundation under
grant PHY90-21984.}}
\smallskip\centerline{\tt gross@puhep1.princeton.edu}
\bigskip
\centerline{and}
\bigskip
\centerline{Andrei Matytsin}
\smallskip\centerline{\tt matytsin@puhep1.princeton.edu}
{\it
\bigskip\centerline{Department of Physics}
\centerline{Joseph Henry Laboratories}
\centerline{Princeton University}
\centerline{Princeton, NJ \ 08544}}
\bigskip

\centerline{\bf Abstract}
\smallskip
\noindent
Large $N$ two-dimensional QCD on a cylinder and on a vertex manifold
(a sphere with three holes) is investigated. The relation between the
saddle-point description and the collective field theory of QCD$_2$ is
established.
Using this relation, it is shown that the Douglas--Kazakov
phase transition on a cylinder is associated with the presence of a gap
in the eigenvalue distributions for Wilson loops. An exact formula for
the phase transition on disc with an arbitrary boundary holonomy is found.
The role of instantons in inducing such transitions is discussed.
The zero-area limit of the partition function on a vertex manifold is
studied. It is found that this partition function vanishes unless
the boundary conditions satisfy a certain selection rule which
is an analogue of momentum conservation in field theory.

\Date{September 1994}

\secno 0
\newsec{Introduction.}

Recently there has been much interest in the study
of QCD$_2$ in the large $N$ limit,
largely motivated by an attempt to find a string representation of
QCD in four dimensions \REFtaylor.  The study of large $N$ QCD$_2$ is
also useful in exploring more genral properties of large $N$ QCD. As such,  a
detailed analysis of the properties of the theory on
the sphere and on the cylinder are interesting.
Perhaps the most important of these properties  is the third order
phase transition for large $N$ QCD on a sphere, discovered
recently by Douglas and Kazakov \REFdouglas.
Physically, this transition is caused by
condensation of instantons. One can demonstrate that the effect of
instantons is negligible when the area of the sphere is small,
and becomes significant as the area reaches the critical value,
$A_{\rm cr}=\pi^2/\lambda$, where $\lambda=g^2 N$ is the large $N$
coupling constant of QCD \REFinst.

The existence of this phase tranition, which is
similar to the phase transition that occurs in the
one-plaquette model \REFgrwit, is of great interest. Such effects
signal a  sharp transition between weak and strong
coupling, between  non-confining
physics and confinement, and between the  a stringy
regime and a non-string regime.
If they were to occur in four dimensional QCD they might be an indication
that the string picture is limited to infrared phenomena
and cannot be analytically
continued to the ultraviolet domain.
One of our goals is to analyse the occurence of such
phase transitions for other geometries.
We will develop methods that allow us to
explore this phase transition in detail.

A phase transition similar to the Douglas-Kazakov transition  occurs in QCD$_2$
on a cylinder with fixed boundary conditions (that is, with fixed
holonomy matrices $U_{C_1}= {\cal P}{\rm exp}\oint_{C_1}
{\cal A}_\mu (x) \thinspace d x^\mu$
and $U_{C_2}= {\cal P}{\rm exp}\oint_{C_2}
{\cal A}_\mu (x) \thinspace d x^\mu$. Here
$C_1$  and $C_2$ are the two circles forming the
boundary of the cylinder).
Obviously, such a phase transition should manifest itself in the
master field of two-dimensional QCD \REFminahan.

A natural set of physical observables in gauge theory is formed by the
Wilson loops $W_n(C)={1\over N}{\rm Tr}\thinspace U^n_C$ with
$U_C= {\cal P}{\rm exp}\oint_C {\cal A}_\mu (x) \thinspace d x^\mu$.
Therefore, we could say that the master field is fully
described by the set of
quantities $W_n(C)$ for all possible contours $C$. Indeed,
$W_n(C)$ satisfy a closed set of equations (the loop equations)
with a well defined large $N$ limit \REFmigrev\REFmami\REFkazkos.
In addition, the expectation
values of Wilson loops do factorize at large $N$. The loop equations,
however, are higly nonlocal and the description of the theory they
provide is very convoluted.

The loop variables are difficult to work with.
For
two-dimensional QCD a better set of variables is readily available.
Indeed, it is known that the two-dimensional QCD is equivalent
to a $c=1$ matrix model with the spatial coordinate compactified
on a circle \REFdou\REFmincone\REFDADDA.
Therefore, the eigenvalue density of the Wilson
matrix $U_C$ will satisfy the appropriate collective field equation
(the Hopf equation) as a function of the area bounded by the
contour $C$. Using this fact we can obtain a formula for Wilson
loops $W_n$ in the large $N$ QCD on a cylinder with arbitrary
boundary conditions.

As a result, the dynamics of Wilson loops on a cylinder
can be desribed by a simple physical picture. Let $\sigma_1(\theta)$
and $\sigma_2(\theta)$,\ $\theta \in [0, 2\pi]$, be the eigenvalue
densities of the boundary holonomy matrices $U_{C_1}$ and $U_{C_2}$.
Consider a one-dimensional compressible fluid living on a circle,
with the equation of state $P=- {\pi^2\over 2}\sigma^2$, $P$ being
the pressure and $\sigma$ the density. Imagine the process
where the fluid moves from an initial configuration, where
its density profile is $\sigma(\theta)=\sigma_1(\theta)$ to a
final one, $\sigma(\theta)=\sigma_2(\theta)$, during the time
interval equal to the total area of the cylinder, $A$. Then
the eigenvalue density for the Wilson matrix $U_C$, where
the contour $C$ cuts the cylinder of total area $A$ into two
cylinders of areas $A_1$ and $A_2$, is just given by the
density of our fluid at the time $A_1$. Consequently,
\eqn\wl{W_n(C)=\int\limits_{0}^{2\pi}\sigma\thinspace(t=A_1, \theta)
\thinspace{\rm e}^{i n \theta}\thinspace d\theta.}

On the other hand, the partition function of the two-dimensional QCD,
at least in the most interesting case of symmetric boundary
conditions $U_{C_1}=U_{C_2}^\dagger$, is dominated by a single
representation of the gauge group. This representation, associated
with its Young tableau, can be characterized by a set of numbers
$h_i= l_i/N$, $i=1, \dots, N$. The density of this set, $\rho_Y(h)$,
plays the crucial role in the analysis of the Douglas-Kazakov
phase transition. However, there is no known equation for determining
$\rho_Y(h)$, except for some special cases.

\ifig\cyllin{The cylinder with boundary contours $C_1$ and $C_2$. The
contour $C_0$, used in the calculation of the dominant Young tableau,
cuts the cylinder in half.}
{\epsfxsize3.5in\epsfbox{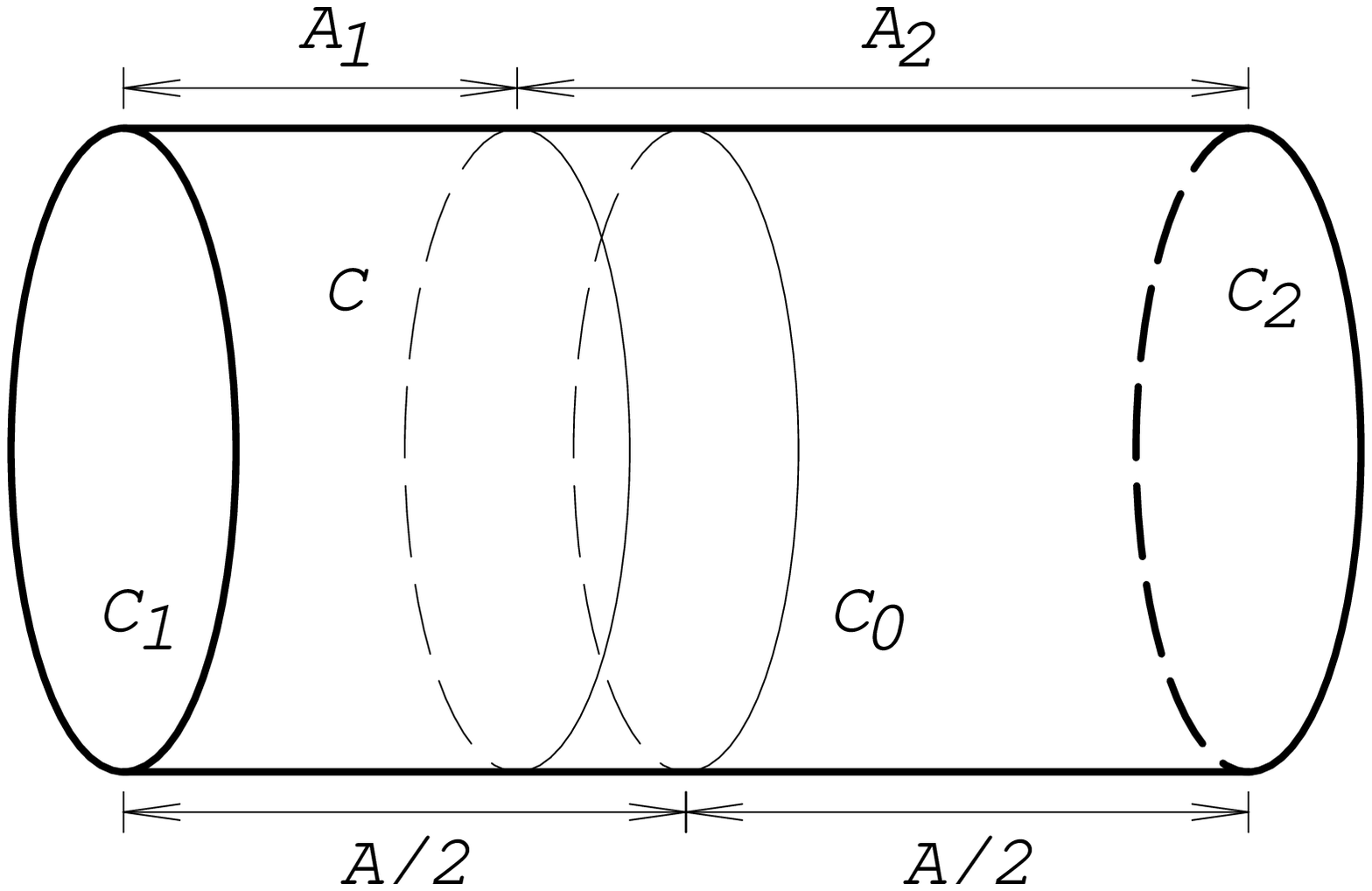}}

This problem shall be solved below, in section 2. We find that, when
$U_{C_1}=U_{C_2}^\dagger$, the Young tableau density $\rho_Y(h)$
is determined by the surprisingly simple formula
\eqn\supsim{\pi\rho_Y\big(-\pi\sigma_0(\theta)\big)=\theta
}
where $\sigma_0(\theta)$ is the density of eigenvalues for the
Wilson matrix $U_{C_0}$. Here the contour $C_0$ cuts the cylinder into
two equal parts, $A_1=A_2=A/2$ (see \cyllin). The density $\sigma_0(\theta)$
can be found using the procedure outlined above\foot{See the
discussion preceding \wl.}. This formula will also allow us
to prove the criterion (formulated by Caselle, D'Adda, Magnea and
Panzeri \REFDADDA)
determining for which boundary conditions the partition function of
${\rm QCD}_2$ can exhibit a Douglas-Kazakov phase transition,
and for which the transition is absent. This criterion applies even in
those situations when no single representation is dominant and
the original method of Douglas and Kazakov leads to a complex
saddle point. Quite remarkably, we discover that for QCD on a disc
(that is, when $\sigma_2(\theta)=\delta(\theta)$)
the transition point can be determined exactly and explicitly to be
$$A_{\rm cr}={\pi\Biggl[ \int {\sigma_1(\theta)\, d\theta\over \pi-\theta}
\Biggr]^{-1}}.$$

As a second application of formula \supsim\ we will consider the
well studied example of QCD on a plane. We find that the structure
of Wilson loops $W_n(A)$ with large values of winding number $n$
experiences qualitative changes as the area enclosed by the loop, $A$,
passes through the critical value $A_{\rm pl}=4$. This phenomenon,
which also can be viewed as a phase transition, is in fact a remnant
of the Douglas-Kazakov transition occuring on a sphere of
finite area $A=\pi^2$.

We will use this result in section 3 to make some general comments about the
implications and meaning of the phase transition.

The collective field theory approach to QCD$_2$ is very powerful, at
least on the cylinder.
The Hopf action, as we shall discuss, propagates Wilson loops,
or eigenvalue densities,
along the cylinder.
\ifig\pantsfig{The vertex manifold with three boundaries,
$C_1$, $C_2$ and $C_3$.}
{\epsfxsize2.75in\epsfbox{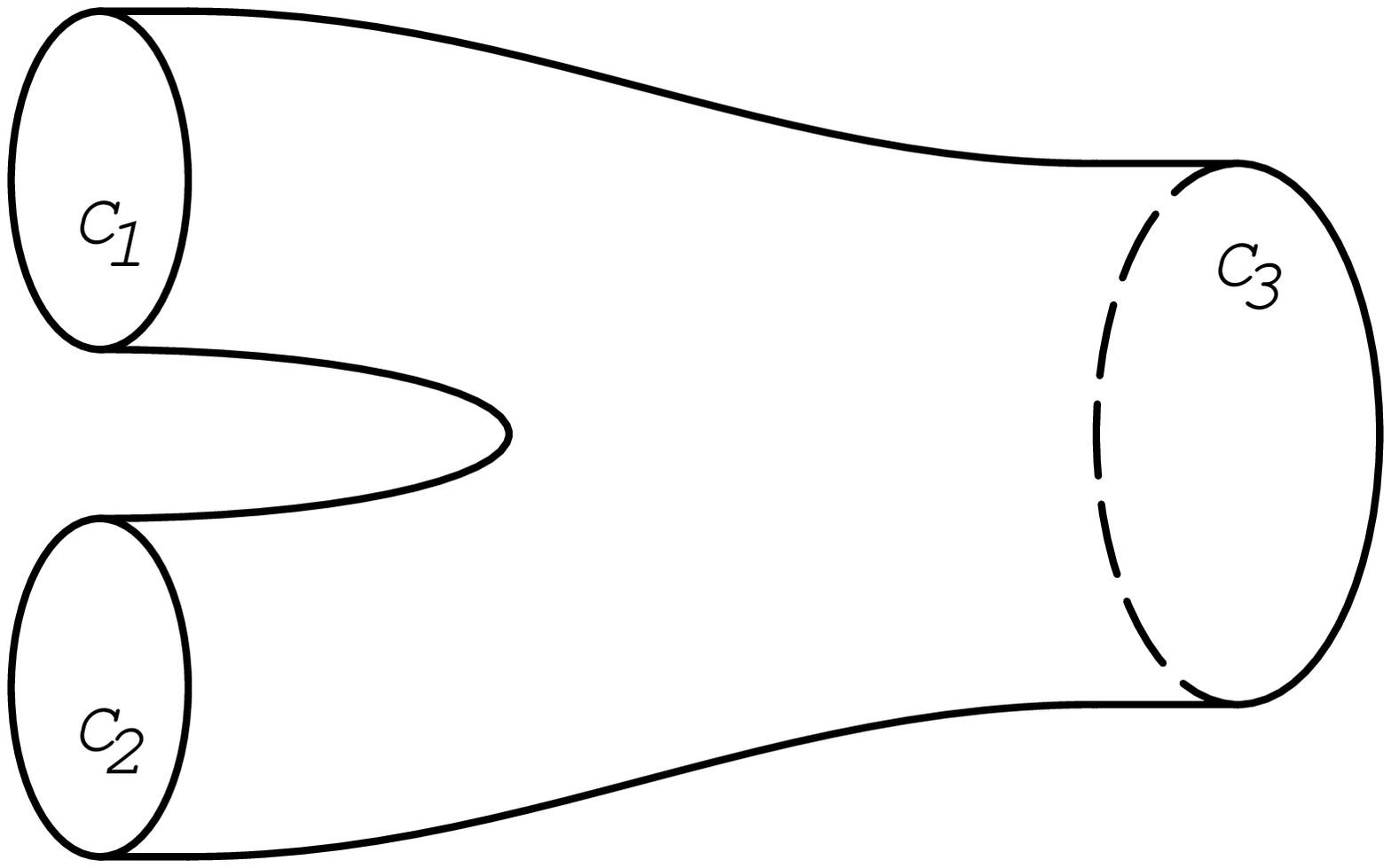}}
To discuss other geometries, in particular surfaces with handles, we have
to know how  to  split the loops at a vertex.
To this end we need the partition function
for QCD$_2$ on a
``pair of trousers"--type manifold (\pantsfig)
(that is, a sphere with three holes).
Moreover, it suffices to consider the limiting case when the
area of this manifold vanishes. Indeed, we can always create the
``pair of trousers" with a finite area by attaching cylinders
to the zero-area ``vertex".

The properties of these string vertices shall be investigated in
section 4. We find that, as $N\to\infty$, the partition function
of the QCD on a ``vertex" vanishes in the zero-area limit, unless
certain selection rule is satisfied. We find and prove this
selection rule and discuss its implications for the string theory
of two-dimensional QCD.

\newsec{Large $N$ QCD on a cylinder.}

The partition function of QCD on a cylinder (fig.1) can be evaluated
exactly for any finite $N$ and equals \REFmigdal\REFrusakov
\eqn\pfcyl{
{\cal Z}_N(U_{C_1}, U_{C_2}|A)=\sum_R \chi_R(U_{C_1})
\chi_R(U_{C_2}^{\dagger})
\thinspace {\rm e}^{-{\lambda A\over 2N}C_2(R)}
}
where $\lambda=g^2 N$ is the large $N$ coupling constant and $A$ the
area of the cylinder.
The summation is over  all irreducible representations $R$
of the gauge group, $\chi_R(U)$ is the character of the matrix $U$
in the representation $R$, and $C_2(R)$ is the quadratic Casimir operator
of this representation.
Since the answer depends only on the product of $\lambda$ and $A$,
we shall set $\lambda=1$.

If the gauge group is $U(N)$, the representations $R$ can be labelled by
a set of integers $+\infty>l_1>l_2>\dots>l_N>-\infty$. Then
\eqn\char{
\chi_R(U)={{\rm det}\big|\!\big|{\rm e}^{i l_j\theta_k}\big|\!\big|\over
J\big({\rm e}^{i\theta_s}\big)}
}
where ${\rm e}^{i \theta_k}; \thinspace\thinspace k=1,\dots, N$
are the eigenvalues of the (unitary) matrix $U$, and
$$J\big({\rm e}^{i\theta_s}\big)=\prod_{j<k}\big({\rm e}^{i\theta_j}-
{\rm e}^{i\theta_k}\big).$$
Also,
$$C_2(R)={N\over 12}(N^2-1)+ \sum_{i=1}^{N}\Bigl(l_i-{N-1\over 2}\Bigr)^2.$$

In this section we will study the partition function \pfcyl\ in
the large $N$ limit. This can be done in two ways. One way  is to
show that for large $N$ \pfcyl\ is dominated by a particular
representation $R$. If this representation is found, it is
possible to express all physical observables in terms of the  indices
$\{l_i\}$ labelling $R$. The second approach is to construct an
appropriate collective field theory describing ${\rm QCD}_2$.
While the latter method is more  physically transparent,
  practical calculations have so far been carried out  using
the first approach. For example, it is not at all
obvious how to find the solution of
the collective field theory which describes the strong coupling phase
of QCD on a sphere.

These two approaches are complementary.
Below we will derive the equation  \supsim\ relating the dominant
representation to the solution of the collective theory. This will allow us
to provide a simple physical interpretation of the formulas for
Wilson loops on a sphere, found recently by Boulatov, Daul and Kazakov
and extend them to the general case of a cylinder.

\subsec{The dominant representation.}

Generally, infinitely many representations contribute to the partition
function \pfcyl. However, as $N\to\infty$ the sum  is dominated by
a single representation. Indeed, for  large $N$ the $U(N)$ characters \char\
behave asymptotically as
\eqn\aschar{
\chi_R(U)\simeq{\rm e}^{N^2 \Xi[\rho_Y(l/N), \sigma(\theta)]
}}
with some finite functional $\Xi[\rho_Y, \sigma]$. In this formula it is
implicit that we take the limit $N\to\infty$ assuming that the
eigenvalue distribution of the unitary $N\times N$ matrix $U$
converges to a smooth function\foot{The eigenvalues of a unitary
matrix lie on the  unit circle in the complex plane and can be
parametrized as $\lambda_j={\rm e}^{i\theta_j}$.}
 $\sigma(\theta)$, $\theta\in [0, 2\pi]$.
In addition, it is assumed that the distribution of parameters
${\tilde y}_i=l_i/N$, which define the representation  $R$, also converges
to another smooth function
$\rho_Y({\tilde y})$, that we can call the
Young tableau  density. The functional
$\Xi$ is, in general, not easy to calculate. However, in some important
cases it can be found explicitly.

As a result,
$$\eqalign{{\cal Z}_N(U_{C_1}, U_{C_2}|A)\simeq &\sum_R
{\rm exp}\biggl[N^2\Bigl\{
\Xi\bigl[\rho_Y({\tilde y}), \sigma_1(\theta)\bigr]+
{\overline {\Xi\bigl[\rho_Y({\tilde y}), \sigma_2(\theta)\bigr]}}\cr &-
{A\over 2}\int \rho_Y({\tilde y}) \Bigl({\tilde y}-{1\over 2}\Bigr)^2
d {\tilde y} -{A\over 24}\Bigr\}\biggr],\cr}
$$
where the bar denotes complex conjugation.
In this formula the summation over all representations $R$ can be thought of
as ``functional integration" over all
possible distributions  of the Young tableaux,
$\rho_Y({\tilde y})$.

For large $N$ we expect that this ``functional
integral" is dominated by a saddle point, which can be determined from the
equation
\eqn\saddpt{{\partial \over \partial {\tilde y}}\biggl[
{\delta
\Xi\bigl[\rho_Y({\tilde y}), \sigma_1(\theta)\bigr]
\over \delta \rho_Y({\tilde y})}
+{\delta
{\overline {\Xi\bigl[\rho_Y({\tilde y}), \sigma_2(\theta)\bigr]}}
\over \delta \rho_Y({\tilde y})}\biggr]=
{A}\Bigl({\tilde y}-{1\over 2}\Bigr).}
Then the free energy of the theory equals
$$\eqalign{F_N(U_{C_1}, U_{C_2}|A)\equiv &\ln {\cal Z}_N(U_{C_1}, U_{C_2}|A)=
N^2\Bigl[\Xi\bigl[\rho_Y(y), \sigma_1(\theta)\bigr]\cr +&
{\overline {\Xi\bigl[\rho_Y(y), \sigma_2(\theta)\bigr]}}-
{A\over 2}\int \rho_Y(y)\thinspace y^2
dy -{A\over 24}\Bigr]+{\cal O}(N^0)\cr}$$
where to simplify the notation we have used a shifted variable
$y={\tilde y}-1/2$. It is also useful to keep in mind that
the derivative of $F_N$ with respect to the area $A$ (the ``specific
heat capacity") is given by\foot{Although $\rho_Y(y)$ by itself
depends on the
area, this dependence does not contribute to the specific heat due to
the saddle point equation \saddpt.}
\eqn\spheat{
{1\over N^2}{\partial F_N\over \partial A}=-{1\over 24}-{1\over 2}\int
\rho_Y(y)\thinspace  y^2\, dy.}

Since the indices $l_i$ are discrete integers,
the differences $l_i-l_{i+1}$ are always
greater than one, and therefore the density $\rho_Y(y)$ must satisfy
the constraint
$\rho_Y(y)\le 1$. The saturation of this bound  causes a phase transition
for large $N$ QCD \REFdouglas.

To see how this phase transition occurs, let
us consider the two-dimensional QCD
on a sphere. We can regard the sphere as a particular case of a cylinder
with boundary conditions $U_{C_1}=U_{C_2}={\rm I}$\ (that
is, $\sigma_1(\theta)=\sigma_2(\theta)=\delta(\theta)$). In this situation
the large $N$ limit of characters is easy to calculate. Indeed,
since
$$\chi_R({\rm I})=d_R=\prod_{i<j}(l_i-l_j),$$
we have
$$\Xi\big[\rho_Y(y), \delta(\theta)\big]=
{1\over 2}\int\rho_Y(y_1)\, \rho_Y(y_2)\, \ln|y_1-y_2|\, dy_1 \, dy_2\, .$$
Then \saddpt\ entails
$${\bf -}\kern-1.1em\int{\rho_Y(u)\, du\over y-u}= {A y \over 2}.$$
The solution of this equation is
\eqn\weasph{
\rho_Y(y)={1\over \pi}\sqrt{A-{A^2 y^2\over 4}}.}

\ifig\densph{The density $\rho_Y(y)$ describing the dominant
Young tableau for a sphere in the weak coupling phase (left)
and in the strong coupling phase (right).}
{\epsfxsize2.25in\epsfbox{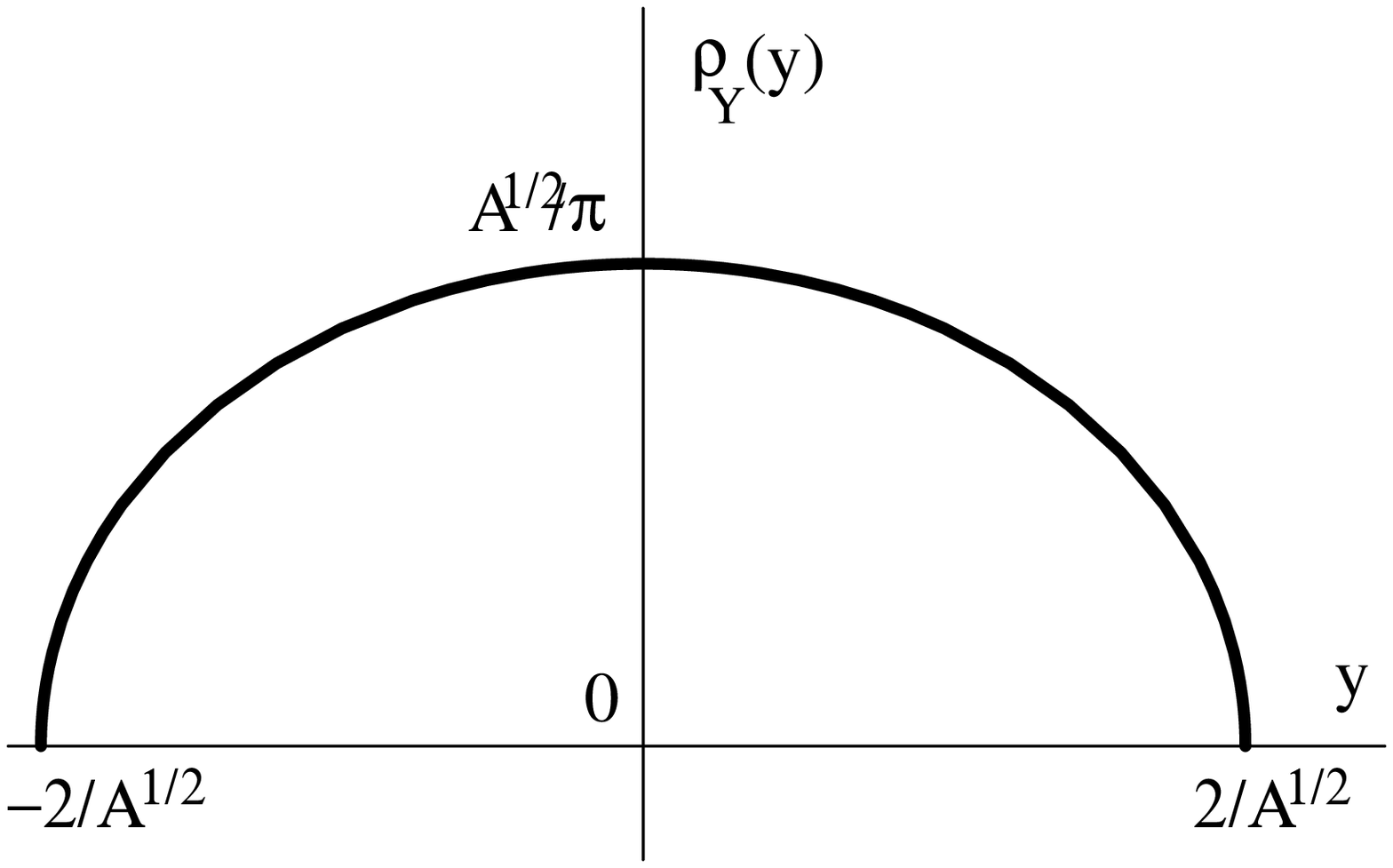}\hskip0.1in
\epsfxsize2.25in\epsfbox{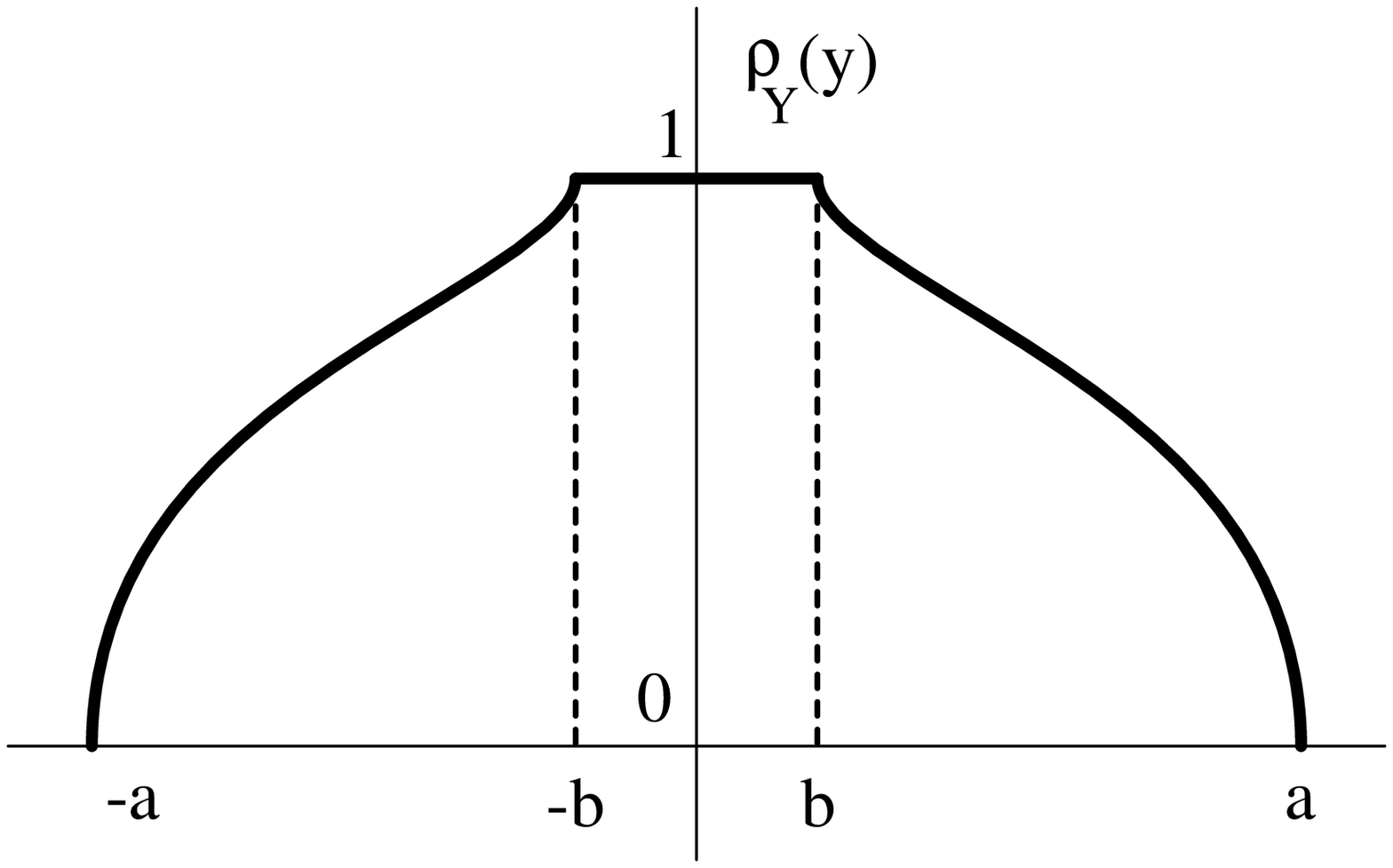}}

However, it obeys the constraint $\rho_Y(y)\le 1$ only if
$A\le \pi^2$. If $A > \pi^2$ then, instead, we must set  $\rho_Y(y)=1$ in some
interval $y\in [ -b, b]$ and  the solution becomes \REFdouglas
\eqn\strsph{
\rho_Y(y)={2\over \pi a y }\sqrt{(a^2-y^2)(y^2-b^2)}\thinspace
\Pi_1\Bigl(-{b^2\over y^2}, {b\over a}\Bigr)}
outside of this interval, for $y\in [ b, a]$.
In this formula
\eqn\ellip{
\Pi_1(x, k)={1\over 2}\int\limits_{-1}^{+1}{du\over 1+x u^2}
{1\over \sqrt{(1- k^2 u^2)(1-u^2)}}}
is the elliptic integral of the third kind and
the constants $b$ and $a$ are fixed by
\eqn\ellpar{
\left\{\eqalign{
&k=b/a\, , \qquad k^{\prime}=\sqrt{1-k^2},\cr
&a\bigl(2{\bf E}(k)- {k^{\prime}}^{2}{\bf K}(k)\bigr)=1,\cr
& a A = 4 {\bf K}(k).\cr
}\right.}
Accordingly, the specific heat \spheat\ is different for $A<\pi^2$
and $A>\pi^2$. In fact, it is possible to show that at the critical
point $A=\pi^2$ the third derivative of the free energy is
discontinuous, indicating a phase transition.

Obviously, a similar transition will occur on the cylinder.
In this case the saddle point equation \saddpt\ becomes very involved.
To find its solution we will have to use an entirely different
treatment of ${\rm QCD}_2$, based on the ideas of the collective field
theory.

\subsec{The collective field theory.}

The large $N$ limit of two-dimensional QCD can  also
be studied with the aid of a collective field theory.
The underlying idea of collective field methods is to
make a change of variables describing the theory so that
the new variables (the ``collective variables")
would have a well defined large $N$ limit \REFjesa.

In our case the partition function ${\cal Z}_N$ depends on
the two sets of eigenvalues $\{\lambda_a^{(1)}=\ee^{i\theta_a^{(1)}}|
a=1, \dots, N \}$
and $\{\lambda_b^{(2)}=\ee^{i\theta_b^{(2)}}|b=1, \dots, N \}$.
The specific  eigenvalues do not tend  to any limit as we take $N$ to
infinity. However, the corresponding eigenvalue distributions
$\sigma_1(\theta)$ and $\sigma_2(\theta)$ do have a large $N$
limit. In fact, $\sigma_1$ and $\sigma_2$ contain all the data
one needs to evaluate the leading $N\to\infty$ asymptotics
of ${\cal Z}_N$.

Therefore, it must be possible to obtain a differential
equation on ${\cal Z}_N$ as a functional of $\sigma_1$ and $\sigma_2$.
To achieve this goal we will write down a differential equation
satisfied by ${\cal Z}_N$ as a function of the $2N$ discrete variables
$\{\theta_a^{(1)}\}$ and $\{\theta_b^{(2)}\}$. Then we will
change variables to $\sigma_1(\theta)$ and $\sigma_2(\theta)$.

As a first step, one represents \pfcyl\ in the form
\eqn\pfexp{\eqalign{
{\cal Z}_N(U_{C_1}, U_{C_2}|&A)={1\over N!}{1\over 2^{N(N-1)}}\cr
\times &\sum_{y_k\in\{\pm {1\over N}, \pm {2\over N}, \dots \}}{{\rm det}
\bigl|\!\bigl|{\rm e}^{i y_a\theta_b^{(1)}}\bigr|\!\bigr|
\thinspace {\rm det}
\bigl|\!\bigl|{\rm e}^{-i y_c\theta_d^{(2)}}\bigr|\!\bigr|
\over \prod\limits_{s<r}\bigl[\sin {\theta_s^{(1)}-\theta_r^{(1)}\over 2}
\sin {\theta_s^{(2)}-\theta_r^{(2)}\over 2}\bigr]}\thinspace
{\rm e}^{-{A\over 2}N\sum\limits_{k=1}^N y_k^2}\cr}
}
with $y_k={1\over N}\bigl(l_k-{N-1\over 2}\bigr)$. It is now obvious
that ${\tilde {\cal Z}}_N\equiv {\cal Z}_N \thinspace
{\rm e}^{{A\over 24}(N^2-1)}$
satisfies the differential equation
\eqn\predas{
2N{\partial\over \partial A}{\tilde {\cal Z}}_N=
{1\over {\cal D}(\theta^{(1)})}\sum_{k=1}^N {\partial^2\over
\partial \theta_k^{(1)2}}\Bigl[ {\cal D}(\theta^{(1)})
{\tilde {\cal Z}}_N\Bigr]
}
where $${\cal D}(\theta^{(1)})\equiv
\prod_{s<r}\sin {\theta_s^{(1)}-\theta_r^{(1)}\over 2}.$$
If ${\tilde {\cal Z}}_N=\exp N^2{\tilde F}_N$, then ${\tilde F}_N$
has a well defined large $N$ limit ${\tilde F}= \lim\limits_{N\to\infty}
{\tilde F}_N$.
The limiting functional ${\tilde F}$ can be determined if
we convert \predas\ into an equation for ${\tilde F}_N$ and then
change  variables from $\{\theta_a^{(1)}\}$ to
$\sigma_1(\theta)$ in that equation.
As a final result of these manipulations, it follows that\foot{The
procedure of changing variables, which is well known but nontrivial,
is described in Appendix A.}
\eqn\actrep{\eqalign{
&{\tilde F}\bigl[\sigma_1(\theta), \sigma_2(\theta)\bigl| A\bigr]=
S\bigl[\sigma_1(\theta), \sigma_2(\theta)\bigl| A\bigr]\cr
&-{1\over 2}\int \sigma_1(\theta)\sigma_1(\varphi)\ln\left|\sin
{\theta-\varphi\over 2}\right|\, d\theta\, d\varphi
-{1\over 2}\int \sigma_2(\theta)\sigma_2(\varphi)\ln\left|\sin
{\theta-\varphi\over 2}\right|\, d\theta\, d\varphi
\cr}}
where the functional $S$ is a solution of
\eqn\hamjac{
{\partial S\over \partial A}={1\over 2}\int\limits_0^{2\pi}
\sigma_1(\theta)\Biggl[\biggl({\partial\over \partial\theta}
{\delta S\over \delta \sigma_1(\theta)}\biggr)^2-{\pi^2\over 3}
\sigma_1^2(\theta)\Biggr].
}
It is helpful to interpret \hamjac\ as the Hamilton--Jacobi
equation for the classical Hamiltonian
\eqn\hamm{
H\bigl[\sigma(\theta), \Pi(\theta)\bigr]={1\over 2}
\int\limits_0^{2\pi}\sigma(\theta)\biggl[\biggl({\partial \Pi
\over \partial \theta}\biggr)^2- {\pi^2\over 3}\sigma^2(\theta)\biggr]
}
with $A$ playing the role of time, the function $\sigma_1(\theta)$
being the canonical
coordinate and $\Pi(\theta)\equiv\delta S/
\delta \sigma_1(\theta)$ the conjugate momentum. The Hamiltonian
\hamm\ is called the Das--Jevicki Hamiltonian \REFjesa.

In fact, the required solution is easy to construct. To do this, we
solve the Hamilton equations of motion
\eqn\hmeq{
{\partial \sigma(\theta)\over \partial t}=
{\delta H[\sigma, \Pi]\over \delta \Pi(\theta)},\qquad
{\partial \Pi(\theta)\over \partial t}=
-{\delta H[\sigma, \Pi]\over \delta\sigma(\theta)}
}
and pick  the particular solution which satisfies the boundary conditions
\eqn\bc{
\left\{\eqalign{\sigma(\theta)\bigr|&_{t=0}=\sigma_1(\theta)\cr
\sigma(\theta)\bigr|&_{t=A}=\sigma_2(\theta)\cr}\right.
}
where, as before, $\sigma_1$ and $\sigma_2$ are the eigenvalue densities of
matrices $U_{C_1}$ and $U_{C_2}$. Then it is possible to prove
that $S[\sigma_1, \sigma_2|A]$
equals the classical action calculated for this particular
solution\foot{If there are several such solutions, the one with
the largest value of $S$ must be chosen.}.

The equations of motion which follow from \hmeq\ are
\eqn\euler{
\left\{\eqalign{&{\partial \sigma\over \partial t}+
{\partial \over \partial \theta}(\sigma v)=0 \cr
&{\partial v\over \partial t}+v{\partial v \over \partial \theta}=
{\partial \over \partial \theta}\left({\pi^2\over 2}\sigma^2\right)
\cr}\right. \, , \quad v(\theta)\equiv {\partial \Pi\over \partial\theta}.
}
These are the Euler equations for a fluid with  negative  pressure
$P=-{\pi^2\over 2}\sigma^2$. The solution we are looking for
corresponds to the process where the density profile $\sigma_1(\theta)$
evolves into the profile $\sigma_2(\theta)$ during a time equal to $A$.

So far we have seen how this formalism gives us a way of calculating
free energy. However, in addition, the collective theory provides
a natural and concise description of Wilson loops in the large $N$
QCD.

The application of the collective field theory to the study of Wilson
loops is based on the following factorization property.
Imagine we cut our cylinder along some contour $C$ (\cyllin) into two pieces
of areas $A_1$ and $A_2$ .
We may ask how the partition function of the whole cylinder
is expressed in terms of the partition functions for these two pieces.

The answer to this question is obvious. Using \pfcyl\ along with
the orthogonality relation for characters
\eqn\orthchar{
\int d U \, \chi_{R_1}(U) \chi_{R_2}(U^{\dagger})=\delta_{R_1, R_2}
}
we easily see that
\eqn\convolu{
{\cal Z}_N(U_{C_1}, U_{C_2}|A)=\int d U_C\thinspace
{\cal Z}_N(U_{C_1}, U_{C}|A_1)\thinspace
{\cal Z}_N(U_{C}, U_{C_2}|A_2).
}
That is to say, to glue the two pieces into a single cylinder
we have to set  their boundary values, at the place we glue,
 equal to the same matrix $U_C$ and then integrate
over all such matrices.
If we treat ${\cal Z}_N(U_{C_1}, U_{C_2}|A)$ as the probability
amplitude for the
process where $U_{C_1}$ evolves into $U_{C_2}$ during the time $A$,  then
\convolu\ is just the convolution property satisfied by the Feynman
path integrals.

However, in the large $N$ limit the integral in \convolu\
is dominated by a single saddle point. Thus the partition
function reduces to the product of the partition functions
of its two pieces. To see this, recall that the characters $\chi_R(U_C)$
depend only on the eigenvalues of $U_C$, and so does ${\cal Z}_N$.
Therefore, the integration in \convolu\ reduces to\foot{The
factor $\bigl[{\cal D}(\theta^{(C)})\bigr]^2$ is the jacobian
which appears when we express the Haar measure $d U_C$ in terms of
eigenvalues.}
$$\eqalign{&{\cal Z}_N(\{\theta_a^{(1)}\}, \{\theta_d^{(2)}\}|A)\cr
&=\int\,  d\theta_1^{(C)}\, \dots \, d\theta_N^{(C)}\,
\bigl[{\cal D}(\theta^{(C)})\bigr]^2\,
{\cal Z}_N\bigl(\{\theta_a^{(1)}\}, \{\theta_b^{(C)}\}\bigr|A_1\bigr)\,
{\cal Z}_N\bigl(\{\theta_b^{(C)}\}, \{\theta_d^{(2)}\}\bigr|A_2\bigr).\cr}$$
Substituting for ${\cal Z}_N$ their expressions in terms of $S$
using \actrep, we obtain
\eqn\acfac{
\ee^{N^2 S[\sigma_1, \sigma_2|A]}=
\int \,  d\theta_1^{(C)}\, \dots \, d\theta_N^{(C)}\,
\ee^{N^2\bigl(S[\sigma_1, \sigma_C|A_1]+S[\sigma_C, \sigma_2|A_2]
\bigr)},
}
where $\sigma_C(\theta)$ is the distribution of angles $\theta^{(C)}_j$
over which the integration is performed.
The integral in \acfac\ is indeed dominated by a saddle
point. We can even find this saddle point exactly from the equation
\eqn\sap{\eqalign{&{\partial\over \partial \theta_j^{(C)}}
\Bigl(S[\sigma_1, \sigma_C|A_1]+S[\sigma_C, \sigma_2|A_2]
\Bigr)\cr &=
{\partial\over \partial \theta}{\delta\over \delta \sigma_C(\theta)}
\Bigl(S[\sigma_1, \sigma_C|A_1]+S[\sigma_C, \sigma_2|A_2]
\Bigr)\Bigr|_{\theta=\theta_j}=0.\cr}
}
Since $S[\sigma_1, \sigma_C|A_1]$ is the action along the
classical trajectory connecting $\sigma_1$ and $\sigma_2$,
the functional derivative of $S$ equals the corresponding canonical
momentum,
$${\partial\over \partial \theta}
{\delta S[\sigma_1, \sigma_C|A_1]
\over \delta \sigma_C(\theta)}=v_1(\theta),$$
where $v_1(\theta)$ is the velocity at the end of the trajectory,
when $\sigma(t, \theta)=\sigma_C(\theta)$.
Similarly,
$${\partial\over \partial \theta}
{\delta S[\sigma_C, \sigma_2|A_2]
\over \delta \sigma_C(\theta)}=-v_2(\theta),$$
with $v_2(\theta)$ being the velocity at the beginning of the
trajectory connecting $\sigma_C(\theta)$ to $\sigma_2(\theta)$.

Hence, the saddle point condition \sap\ states that $v_1(\theta)=
v_2(\theta)$. That is to say, the two trajectories $\sigma_1\to
\sigma_C$ and $\sigma_C\to \sigma_2$ can be joined so that
the velocity is continuous. The resulting compound trajectory
$\sigma_*(t, \theta)$ is just the solution of the collective field
equations \euler\ which describes the original cylinder
of area $A$ with the boundary conditions $\sigma_1$ and $\sigma_2$.
To find this trajectory, we must solve \euler\ together with the
boundary conditions \bc. Then
the saddle point density $\sigma_C(\theta)$ can be determined
as $\sigma_*(t=A_1,\theta)$.

Using these arguments we can immediately evaluate any Wilson loop
average on a cylinder. In general, a Wilson loop with the
winding number $n$ around the contour $C$ equals
\eqn\willoop{\eqalign{
W_n(C)&\equiv \Bigl\langle{1\over N} {\rm Tr}\,U_C^n\Bigr\rangle\cr=&
{1\over {\cal Z}_N(U_{C_1}, U_{C_2}|A)}\int dU_C\thinspace
{\cal Z}_N(U_{C_1}, U_{C}|A_1)\, \Bigl[{1\over N} {\rm Tr}\,U_C^n\Bigr]\,
{\cal Z}_N(U_{C}, U_{C_2}|A_2).\cr}}
At large $N$ the integral in \willoop\ is again
dominated by a saddle point.
Moreover, this is exactly the same saddle point which dominates the
integral \acfac. Indeed, the new term in the integrand,
$(1/N){\rm Tr}\,U_C^n=(1/N)\sum_{j=1}^N\theta_j^n$, is finite
as $N\to\infty$ and does not affect the
position of the saddle point, which is determined by the balance of
exponentially large, $\ee^{N^2 S}$, terms. As a result, we can
calculate $W_n(C)$ at large $N$ evaluating $(1/N)\sum_{j=1}^N\theta_j^n$
at the saddle point, to obtain
\eqn\wla{
W_n(C)=\int \limits_0^{2\pi} \sigma_*(t=A_1,\theta)\,
\ee^{i n \theta}\, d\theta.}
We have thus  proved that the solutions of the
collective field equations have the physical meaning of yielding the
dominant eigenvalue distributions which, upon Fourier
transform, produce the Wilson loops\foot{A different
formula for Wilson loops on a sphere has been obtained by Boulatov, Daul
and Kazakov \REFdaulkaz. Using the duality relation \supsim\ it is possible
to prove that it is equivalent to \wla.} for  large $N$ QCD.

\subsec{The duality relation for QCD on a cylinder.}

Now we are in a position to find explicitly the representation
which dominates the partition function. Let us first consider
the symmetric case $U_{C_1}=U_{C_2}$. Later it will be obvious that
our treatment easily gengeralizes to include all other boundary conditions.
We will show that $\rho_Y(y)$ satisfies the equation
\eqn\ssupsim{\pi\rho_Y\big(-\pi\sigma_0(\theta)\big)=\theta
}
where $\sigma_0(\theta)$ is the solution of collective equations
\euler\ taken at the middle of the cylinder, for $t=A/2$.

To illustrate this relation let us reexamine  the example of
QCD on a sphere \REFDADDA. There $\sigma_1(\theta)=\sigma_2(\theta)=
\delta(\theta)$ and the solution of the collective field problem
\bc, \euler\ is given by the self-similar evolution of a semicircular
eigenvalue distribution
\eqn\sistar{
\sigma_*(t, \theta)={1\over \pi}\sqrt{\mu(t)- {\mu^2(t)
\theta^2\over 4}}}
where the scale $\mu(t)$ changes in time according to
\eqn\muoft{\mu(t)={A\over t(A-t)}.}
Indeed, if we plug the ansatz \sistar~  into   \euler~
we find that
\eqn\sol{
v(t, \theta)=\alpha(t)\theta, \ {\dot \mu}+2 \alpha\mu=0, \
{\dot \alpha}+\alpha^2+{\mu^2\over 4}=0
}
 whose solution is given by \muoft .
At $t=A/2$ we have $\mu(t)=\mu_0=4/A$ and
$$\sigma_0(\theta)=\sigma_*\Bigl(t={A\over 2}, \theta\Bigr)=
{2\over \pi A}\sqrt{A-\theta^2}.$$
Using \weasph\ we can check that, indeed,
$\pi\rho_Y\big(-\pi\sigma_0(\theta)\big)=\theta$
is satisfied.

In general, the solution of collective field equations cannot be
written down in an explicit form. However, these equations themselves
can be integrated thus reducing the problem to a single implicit
equation on $\sigma_*(t, \theta)$. To see this, let us introduce
a new unknown function
\eqn\hop{
f(t, \theta)=v(t, \theta)+i\pi \rho(t, \theta).}
Then the collective field equations \euler\ can be reduced to
a single equation
\eqn\hopf{
{\partial f\over \partial t}+f{\partial f\over \partial\theta}=0}
known as the Hopf (or Burgers) equation.
Its solution can be found from the implicit formula\foot{Since
$\theta$ is a coordinate
on a circle, $f(t, \theta)$ and $f_0(\theta)$
should be regarded as periodic functions of $\theta$  in this formula .}
\eqn\hopfsol{
f(t, \theta)=f_0\bigl(\theta - t \, f(t, \theta)\bigr)}
where the function $f_0(\theta)$ represents the initial data,
$f_0(\theta)\equiv f(t=0, \theta)$.

We can use \hopfsol\ to relate $\sigma_0(\theta)$ to the
density at the boundary, $\sigma_1(\theta)$. Indeed, if the
boundary densities are the same ($\sigma_1(\theta)=
\sigma_2(\theta)$) then, by symmetry, the velocity at
$t=A/2$ vanishes:
$$v\Bigl(t={A\over 2}, \theta\Bigr)=0.$$
Thus we can use
$$f_0(\theta)=f\Bigl(t={A\over 2}, \theta\Bigr)=i\pi\sigma_0(\theta)$$
as an initial condition in \hopfsol. After the time $\Delta t=A/2$
this $f_0$ should evolve into
$$f(t=A, \theta)=v_1(\theta)+i\pi\sigma_1(\theta),$$
with $\sigma(t=A, \theta)=\sigma_1(\theta)$
and some $v_1(\theta)=v(t=A, \theta)$.
Thus we obtain the equation
\eqn\siocon{
i\pi\sigma_0\Bigl[\theta-{A\over 2}\bigl(v_1(\theta)
+i\pi\sigma_1(\theta)\bigr)\Bigr]=
v_1(\theta)+i\pi\sigma_1(\theta)}
constraining $\sigma_0(\theta)$ (and also $v_1(\theta)$) as soon as
$\sigma_1(\theta)$ is known. Although it would be very difficult to
express $\sigma_0$ through $\sigma_1$ in an explicit form,
\siocon\ contains just what we need to solve for the dominant
Young tableau densiity $\rho_Y(y)$ in \saddpt.

To do it we will have to evaluate the functional derivatives of
$\Xi[\rho_Y, \sigma_1]$ in \saddpt.
This can be done if we observe that the $U(N)$ characters \char\
can be represented as analytic continuations of the Itzykson--Zuber
integral \REFizube
\eqn\itzyk{I_N(A, B)\equiv \int dU\, \ee^{N{\rm Tr}(AUBU^{\dagger})}=
{{\rm det}\, \bigl|\!\bigr|\ee^{N a_k b_j}\bigl|\!\bigr|\over
\Delta({a_k})\Delta(b_j)}.}
In this formula $A$ and $B$ are arbitrary hermitian matrices,
$a_k$ and $b_j$ are their eigenvalues and $\Delta(a_k)\equiv
\prod_{i<j}(x_i-x_j)$ is the Van der Monde determinant.
Setting $a_k=l_k$, $b_j=\theta_j$ and analytically continuing
$a_k\to i a_k$, we see that
\eqn\deter{
{{\rm det}\, \bigl|\!\bigr|\ee^{N a_k b_j}\bigl|\!\bigr|}
\rightarrow J\big({\rm e}^{i\theta_s}\big)\, \chi_R(U).}

Therefore, we can use  the known expressions for the large $N$
limit of the Itzykson--Zuber integral \REFmat\ to find the functional
$\Xi$ in \aschar. In particular, if  as $N\to\infty$   the
distributions of $\{a_k\}$ and $\{b_j\}$ converge to smooth
functions $\alpha(a)$ and $\beta(b)$, then asymptotically
\eqn\izas{\eqalign{
&I_N(A, B)\simeq\exp N^2\biggl\{
S[\alpha, \beta]
+{1\over 2}\int \alpha(a)\,  a^2 \, da +{1\over 2}\int\beta(b) \, b^2
\, db\cr
&- {1\over 2}\int \alpha(a)\,\alpha(a^{\prime})
\, \ln|a-a^{\prime}|\, da\, da^{\prime}-
{1\over 2}\int \beta(b) \, \beta(b^{\prime})\,
\ln|b- b^{\prime}|\, db\, db^{\prime}\biggr\}\cr}}
with some functional $S$ depending on the two distributions
$\alpha$ and $\beta$.

Although there is an explicit expression for
$S[\alpha, \beta]$, we need to know only its functional derivative
\eqn\fderone{
{\tilde U}(a)={\partial \over \partial a}{\delta\over \delta \alpha(a)}
S[\alpha, \beta].}
To determine it we construct  a pair of functions
$\bigl\{G_+(z), G_-(z)\bigr\}$ such that
\eqn\pairf{
\left\{\eqalign{
&G_+\bigl(G_-(z)\bigr)=G_-\bigl(G_+(z)\bigr)=z,\cr
&{\rm Im}\, G_+(a+i 0)=\pi \alpha(a), \cr
&{\rm Im}\, G_-(b+i 0)=-\pi\beta(b).\cr
}\right.}
Then
\eqn\resutilde{
{\tilde U}(a)={\rm Re}\, G_+(a+i 0) - a,}
and, in addition,
\eqn\resvtilde{
{\tilde V}(b)\equiv {\partial \over \partial b}
{\delta\over \delta \beta(b)}S[\alpha, \beta]=
- {\rm Re}\, G_-(b+i 0) + b.}

If we want to apply these formulas to the calculation of
characters, we must be able to perform an analytic continuation
$l_k\to i l_k$. This can be done by introducing an additional
parameter  $t$  and rescaling  $l_k\to t l_k$. We will keep $t$
in our formulas up to the very end of the calculations, where
we will set $t=i$.

In this case the density of points $y_k(t)={l_k(t)\over N}
={l_k\over Nt}$  equals\foot{From now on, we omit the subscript
``$Y$" in $\rho_Y(y)$.}
\eqn\rhot{
\rho_t(y)={1\over t}\, \rho\Bigl({y\over t}\Bigr).}
Therefore, we must distinguish between the derivatives with
respect to $\rho(y)$ and $\rho_t(y)$.
The quantity which enters the master field equation \saddpt\
is the derivative with respect to $\rho(y)$,
\eqn\derrhoy{
U(y)={\partial \over \partial y}
{\delta\over \delta \rho(y)}S\bigl[\rho_t(y), \sigma_1(\theta)\bigr].}
On the other hand, the constraints \pairf\ will yield not $U(y)$, but
\eqn\derrhot{
U_t(y)={\partial \over \partial y}
{\delta\over \delta \rho_t(y)}S\bigl[\rho_t(y), \sigma_1(\theta)\bigr].}
However, these two derivatives are connected by a simple relation
\eqn\sirel{U(y)=t\,  U_t(ty).}
Indeed, if
$$H_t(y)={\delta\over \delta \rho_t(y)}
S\bigl[\rho_t(y), \sigma_1(\theta)\bigr]$$
then
$$\eqalign{&{\delta S\bigl[\rho_t(y), \sigma_1(\theta)\bigr]
\over \delta \rho(y)}= {\delta\over \delta \rho(y)}
\int dz\, \delta\rho_t(z)\, H_t(z)\cr
&={\delta\over \delta \rho(y)}\int dw\, \delta\rho(w)\, H_t(tw)=
H_t(ty)\cr}$$
where we made the change of variables $w=z/t$ and utilized
\rhot. Now it is easy to see that
$$U(y)={\partial \over \partial y}H_t(ty)=t\,  U_t(ty).$$

Therefore, using \aschar, \itzyk, \deter, \izas\
and taking into account that $\ln J(\ee^{i\theta_s})$
does not contribute to the derivative with respect to
$\rho(y)$, we obtain
\eqn\xider{\eqalign{&
{\partial \over \partial y}
{\delta \Xi\bigl[\rho(y), \sigma_1(\theta)\bigr]\over \delta \rho(y)}=
{\partial \over \partial y}{\delta\over \delta \rho(y)}
\biggl\{S\bigl[\rho_t(y), \sigma_1(\theta)\bigr]+{1\over 2}
\int \rho_t(y)\, y^2\, dy\biggr\}\biggr|_{t=i}\cr
&=\Bigl[U(y) +t^2 y\Bigr]_{t=i}=\Bigl[t\, U_t(ty)\Bigr]_{t=i} -y.\cr}}

Now we are able to produce the solution of \saddpt. If $\sigma_1=\sigma_2$,
this equation takes the form\foot{We remember that
$y={\tilde y}-{1\over 2}$.}
\eqn\sadt{{\rm Re}\, {\partial \over \partial y}
{\delta \Xi\bigl[\rho(y), \sigma_1(\theta)\bigr]\over \delta \rho(y)}
={A\over 2}y.}
We will prove that if $\pi\rho(y)$ is the inverse function of
$\pi\sigma_0(\theta)$ then \sadt\ is indeed true.

To this effect we exibit a pair of functions $\bigl\{G_+(z), G_-(z)\bigr\}$
which satisfy the constraint \pairf\ for
$\alpha(a)=\rho_t(a)$ and $\beta(b)=\sigma_1(b)$:
\eqna\cosol
$$\eqalignno{
&G_+(z)= -{A\over 2 }z +i\pi\rho_t(z), &\cosol a\cr
&G_-(z)= -v_1(\theta)-i\pi\sigma_1(\theta)&\cosol b\cr
}$$
where $v_1$ is the velocity appearing in \siocon. This would
imply that, according to \resutilde,
$$U_t(y)={\rm Re}\, G_+(y+i0)-y=-{A\over 2} y -y$$
and using \sirel\ we immediately see that \sadt\ is satisfied.

The conditions \pairf\ are easy to check. While the second and the
third condition in \pairf\ are satisfied by construction,
the first one follows from \siocon. Indeed,
we can rewrite \siocon\ as
$$\pi\sigma_0\Bigl[\theta+{A\over 2}G_-(\theta)\Bigr]=
i G_-(\theta).$$
Using $\pi\rho\bigl(-\pi\sigma_0(\theta)\bigr)=\theta$
and $\rho_t(y)|_{t=i}={1\over t}\rho({y\over t})|_{t=i}=-i\rho(-i y)$,
we deduce
$$\theta+{A\over 2}G_-(\theta)=\pi\rho\bigl(-i G_-(\theta)\bigr)=
i\pi\rho_t\bigl(G_-(\theta)\bigr).$$
Thus
$$-{A\over 2}G_-(\theta)+i\pi\rho_t\bigl(G_-(\theta)\bigr)=\theta$$
which, together with \cosol{a}\ proves
$G_+\bigl(G_-(z)\bigr)=z$, completing our
construction.

If the two boundary densities $\sigma_1$ and $\sigma_2$ are not
identical then the partition function may or may not be dominated
by a single representation. These two situations can be distinguished
by looking at the collective field trajectory connecting $\sigma_1$
to $\sigma_2$. If at some moment $t_0$ the velocity $v(t_0, \theta)$
vanishes simultaneously for all $\theta\in [0, 2\pi]$, then the
dominant representation exists and is determined by\foot{Now $t_0$
does not have to equal $A/2$.}
$$\pi\rho \bigl(-\pi \sigma(t_0, \theta)\bigr)=\theta.$$
On the other hand, if this is not the case the saddle point determined
by \saddpt\ shifts to the complex domain and no single representation
dominates ${\cal Z}[U_{C_1}, U_{C_2}|A]$.

\subsec{The Douglas--Kazakov phase transition on a cylinder.}

A  phase transition similar to the Douglas--Kazakov phase transition
on a sphere occurs in the large $N$ QCD on a  cylinder.

Technically, this phase transition takes place when the range of
values of the function $\pi \rho_Y(y)$ fills the whole interval
$[0, 1]$. Then in the strong coupling phase there are continuous
intervals of $y$ where $\rho_Y(y)\equiv 1$, while in the weak
coupling phase everywhere $\rho_Y(y)<1$.

Using the duality relation \ssupsim\ we can prove a more physical
criterion which determines the transition point. Indeed, since
$\pi \rho_Y(y)$ and $\pi \sigma_0(\theta)$ are inverse functions\foot{The
functional inversion in this context should be  carefully understood
since, strictly speaking, the inverse of $\pi \rho_Y(y)$
has several branches which should be chosen in an appropriate way.
As a consequence, the domain of $\pi \sigma_0$ is $[-\pi, \pi]$ and
not $[0, \pi]$.} the range of values of $\pi \rho_Y(y)$ coincides with the
domain of definition for $\pi \sigma_0(\theta)$.
Therefore at the point of phase transition the domain of $\pi \sigma_0$
ranges from $-\pi$ to $\pi$ thus filling the whole circle.
On the contrary, in the weak coupling phase this domain is restricted
to some interval inside a circle and there is a gap where
$\sigma_0(\theta)$ vanishes.  Thus this phase transition is precisely
of the form found originally for the  unitary matrix model \REFgrwit .

Let us now look at the whole collective field trajectory $\sigma_*(t,
\theta)$. As we have seen from \euler,  $\sigma_*$ evolves as the
density of the fluid with negative pressure. Since such fluid tends
to collapse the moment of time when its velocity vanishes corresponds to
the maximum expansion of the fluid. The density profile at this moment
is, by definition, $\sigma_0(\theta)$. Therefore, if the support of
$\sigma_0$ has a gap, this gap can only increase during further evolution,
and the support of $\sigma_*(t, \theta)$ at any other $t$ will have a gap
as well. As a result, in the weak coupling phase the collective density
exibits the gap at all times, while in the strong coupling phase there
are time intervals when the gap disappears. This provides a criterion
for the Douglas--Kazakov phase transition which is more general than
the condition $\rho_Y(y)=1$. Indeed, it is applicable even
when no single representation dominates the partition function and
no $\rho_Y(y)$ exists.

\ifig\sphpicture{The Young tableau
density $\rho_Y(y)$ (left) and the collective
field density $\sigma_0(\theta)$ (right) for the QCD on a sphere.
In accordance with the duality relation the plots of $\pi\rho_Y(y)$
and $\pi\sigma_0(\theta)$ can be obtained from each other by
a reflection with respect to the axis $x=y$.}
{\epsfxsize2.2in\epsfbox{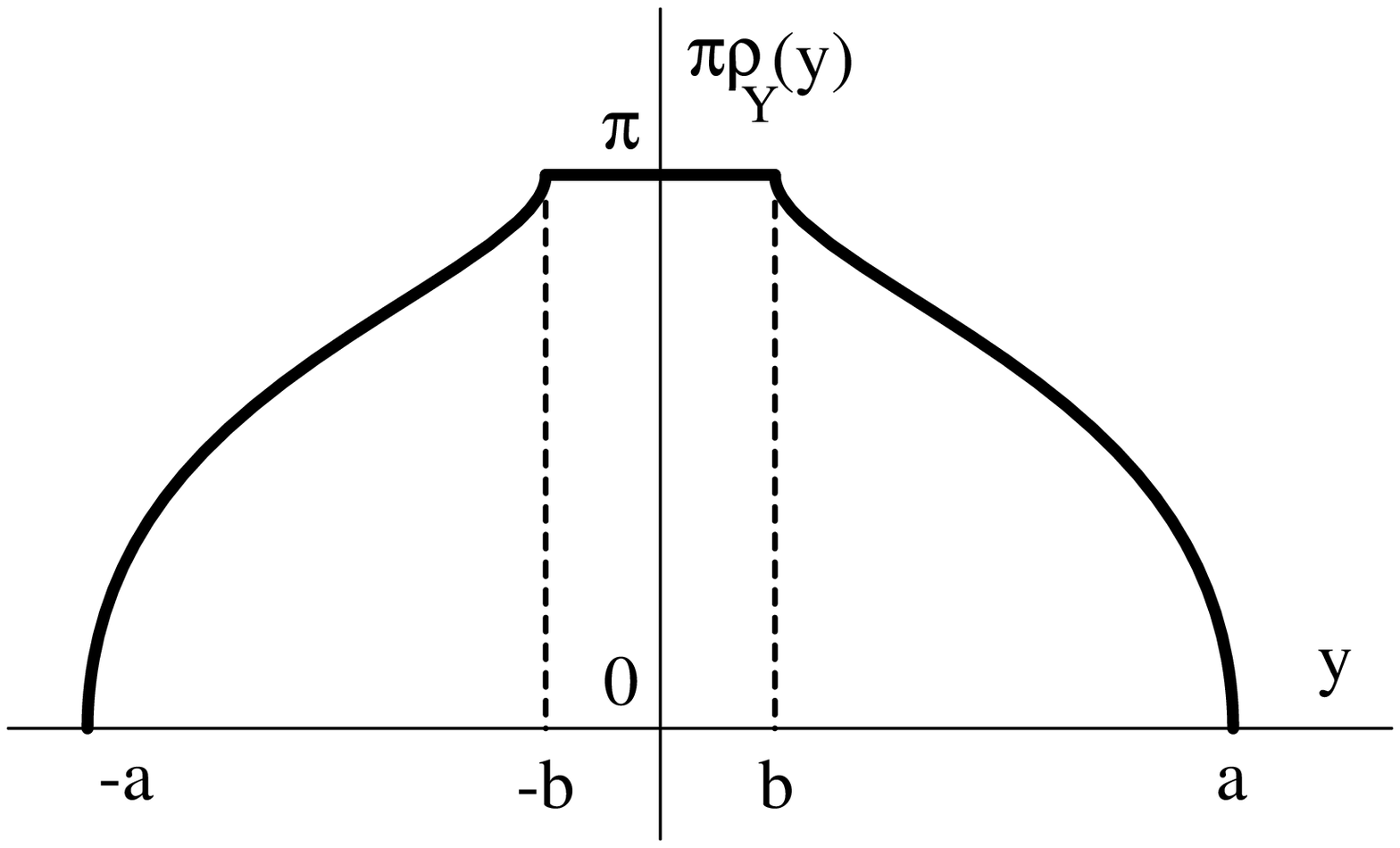}\hskip0.1in
\epsfxsize2.2in\epsfbox{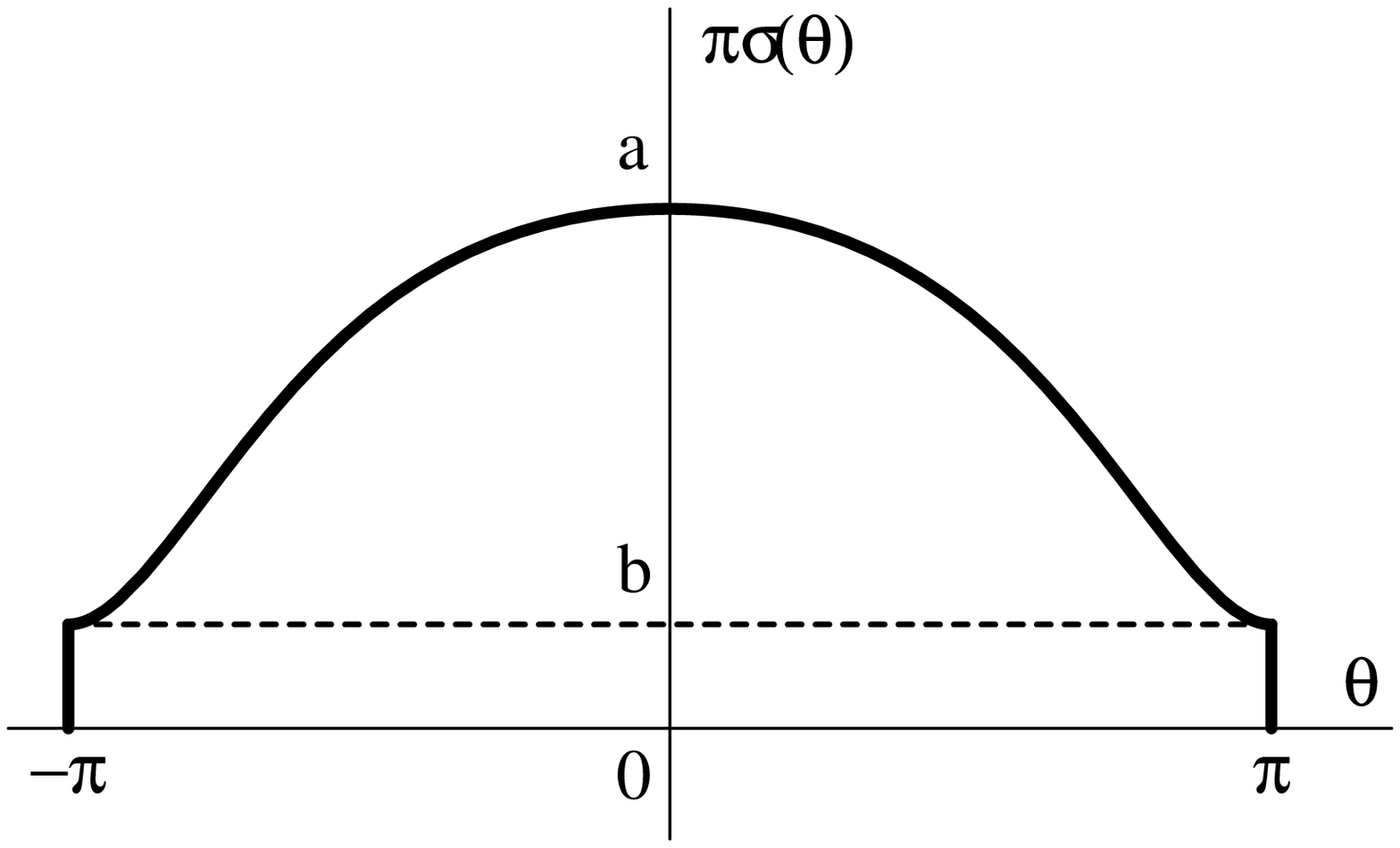}}

Returning to the QCD on a sphere, we can now construct the collective
field trajectory in the strong coupling phase\foot{The trajectory in
the weak coupling phase is given by \sistar.}.
We are looking for a solution of the Hopf equation \hopf\
with the boundary conditions ${\rm Im}\, f(t=0, \theta)=
{\rm Im}\, f(t=A, \theta)=\pi \delta(\theta)$. Surprisingly,
the solution of this problem is quite complicated. Indeed, by the
duality formula \ssupsim\ the value of $f(t=A/2, \theta)=
\pi\sigma_0(\theta)$ is the inverse function of $\pi\rho_Y(y)$
which, in turn, is given by the elliptic integral \strsph.
Since this solution describes the strong coupling phase
the support of $\sigma_0$ has no gap (\sphpicture).

However, the gap will inevitably appear as $\sigma_0(\theta)$
evolves according to the Hopf equation. After all, the final
result of this evolution is the density $\sigma_1(\theta)=
\delta(\theta)$ with support consisting of a single point.

Since our boundary conditions are even, so is $\sigma_*(t, \theta)$
for any $t$. Therefore, the gap in the support of $\sigma_*$ will
be centered around $\theta=\pi$. In particular, we can find the moment
of time when this gap just appears by looking at the solutions of
$\sigma_*(t, \theta=\pi)=0$.  We will find that this equation
has solutions only if $t\in [0, \, {1\over 2}A-\tau_0(A)]\cup[{1\over 2}A+
\tau_0(A), \, A]$. On the other hand, when $t\in ]{1\over 2}A-\tau_0(A),
\, {1\over 2}A+\tau_0(A)[$ the gap is absent and $\sigma_*(t, \theta)
\ne 0$.

Since the Wilson loops are nothing but the Fourier coefficients
of $\sigma_*(t, \theta)$, we conclude that the structure of small
loops (with area less than the \lq\lq critical" value $t_{\rm cr}=
{1\over 2}A-\tau_0(A)$) resembles the structure of loops in the weak coupling
phase, even though we are in the strong coupling phase.
On the other hand the loops with larger area are truly characteristic
of the strong coupling phase. Physically, we can distinguish between
these two types of loops by considering their behavior as the winding
number $n$ goes to infinity. Then at fixed area of the contour
the strong coupling type of a loop\foot{By definition,
this is the loop for which the corresponding collective field
density has no gap. Alternatively, the weak coupling type
means that the gap is present. This should not be confused
with the strong and weak coupling phases of the Douglas--Kazakov
transition.} decreases exponentially with $n$
\eqn\expdec{
W_n(C)={\cal O}(\ee^{-\alpha(C) n}), \quad \alpha(C)>0
}
while the weak coupling type of a loop decreases at most
algebraically, as ${\cal O}(n^{-p})$. The same is true for
the general case of a cylinder.

Continuing our example, let us find
how the critical Wilson loop area depends on $A$. Obviously,
$\tau_0(A)$ vanishes at the Douglas--Kazakov transition, when
$A=\pi^2$, and then grows as $A$ increases.

The evolution of $\sigma_*(t, \theta=\pi)$ is easy to study. We
can use \hopfsol\ with the initial condition $f_0(\theta)=
i\pi\sigma_0(\theta)$ to obtain\foot{Since we set the initial
condition at $t_0=A/2$, the time of evolution in this equation
is $t-{1\over 2}A$, not $t$.}
$$i\pi\sigma_0\Bigl[\theta- \Bigl(t-{A\over 2}\Bigr)f(t, \theta)\Bigr]
=f(t, \theta).$$
However, due to the symmetry $\theta\to -\theta$ the velocity
$v(t, \theta)$ vanishes at $\theta=0$ and $\theta=\pi$.
Thus $f(t, \pi)=i\pi\sigma_*(t,\pi)$ and
$$\pi\sigma_0\Bigl[\pi -i \Bigl(t-{A\over 2}\Bigr)\pi \sigma_*(t,\pi)
\Bigr]=\pi \sigma_*(t,\pi).$$
Finally, using the duality \ssupsim\ we obtain an equation for
$z(t)=\pi \sigma_*(t,\pi)$,
\eqn\eqz{
\pi - i \Bigl(t-{A\over 2}\Bigr) z(t)= \pi\rho_Y\bigl(z(t)\bigr)}
with $\rho_Y(y)$ given by \strsph.

The solution $z(t)$ we are looking for is a real function of $t$.
Moreover,  at all times $0\le z(t)< b$. Indeed, at $t=t_0=A/2$
the expansion of our hypothethic fluid is maximal and so is its
density at $\theta=\pi$. When the fluid collapses towards
$\theta=0$, $\sigma(t, \pi)$ decreases to zero.

Mathematically, if $z\in [0, b]$
\eqn\rhoim{
\pi\rho_Y(z)=i{2\over a z}\sqrt{(a^2-z^2)(b^2-z^2)}\> \Pi_1\Bigl(-
{b^2\over z^2}, k\Bigr).}
Since in this case $x=-{b^2\over z^2}<-1$, the elliptic integral
$\Pi_1(x, k)$ develops an imaginary part. This happens because
the poles of the integrand at $t^2=-{1\over x}$ are now located
within the integration region. Using the rule ${1\over x}=
{\cal P}{1\over x}-i\pi \delta(x)$, we get
\eqn\ellim{
\Pi_1(x, k)={1\over 2}-\kern-1.15em\int\limits_{-1}^{+1}{du\over 1+x u^2}
{1\over \sqrt{(1- k^2 u^2)(1-u^2)}}
-{i\pi \over \sqrt{|x|\bigl(1+{k^2\over x}\bigr)
\bigl(1+{1\over x}\bigr)}}.}
Substituting this into \rhoim\ and using $k=b/a$, we see that
$$\pi\rho_Y(z) =\pi +{i\over a z}\sqrt{(a^2-z^2)(b^2-z^2)}
-\kern-1.135em\int\limits_{-1}^{+1}{du\over
(1+x u^2)\sqrt{(1- k^2 u^2)(1-u^2)}}.$$
Therefore \eqz\ reduces to a single real equation
\eqn\sire{
\Bigl(t-{A\over 2}\Bigr)z=-{1\over a z}
\sqrt{(a^2-z^2)(b^2-z^2)}
-\kern-1.15em\int\limits_{-1}^{+1}{du\over
(1-u^2{b^2\over z^2})\sqrt{(1- k^2 u^2)(1-u^2)}}.}

It is easy to see that $z=0$ is always a solution of this
equation. However, for certain $t$ it has another solution
which can be determined from
$$t-{A\over 2}=-{1\over a}\sqrt{(a^2-z^2)(b^2-z^2)}
-\kern-1.15em\int\limits_{-1}^{+1}{du\over
(z^2- u^2 b^2)\sqrt{(1- k^2 u^2)(1-u^2)}}.$$
Such $z(t)$ decreases as $t-{A\over 2}$ deviates from
zero, so that when $t$ goes outside of the interval
$]{1\over 2}A-\tau_0(A),\, {1\over 2}A+\tau_0(A)[$
this solution ceases to exist. Therefore, at
$t={1\over 2}A +\tau_0(A)$ we have $z(t)=0$ and
$$\tau_0(A)=-b \lim\limits_{z\to 0}
-\kern-1.10em\int\limits_{-1}^{+1}{du\over
(z^2- u^2 b^2)\sqrt{(1- k^2 u^2)(1-u^2)}}.$$
Using the $x\to 0$ asymptotics
\eqn\asymp{
-\kern-1.105em\int\limits_{-1}^{+1}{du\over
(u-x)\sqrt{(1- k^2 u^2)(1-u^2)}}=
2 k^2 x \int\limits_{0}^{1}{u^2 \, du\over
\sqrt{(1- k^2 u^2)(1-u^2)}} +{\cal O}(x^2)}
we conclude that
$$\lim\limits_{z\to 0}-\kern-1.108em\int\limits_{-1}^{+1}{du\over
(z^2- u^2 b^2)\sqrt{(1- k^2 u^2)(1-u^2)}}=
-{2k^2\over b}\int\limits_{0}^{1}{u^2 \, du\over
\sqrt{(1- k^2 u^2)(1-u^2)}}$$
and finally
\eqn\tauzero{
\tau_0(A)={1\over ab}\int\limits_{-b}^{b}{\lambda^2\, d\lambda
\over \sqrt{(a^2- \lambda^2)(b^2-\lambda^2)}}
={2\over b}\bigl({\bf K}(k)- {\bf E}(k)\bigr).}

\ifig\tcraplot{The critical Wilson loop area $t_{\rm cr}$ as
a function of the size of the sphere. As $A\to\infty$ (this
corresponds to the QCD on a plane)
$t_{\rm cr}$ converges to a finite limit $t_{\rm cr}=4$.}
{\epsfxsize3.5in\epsfbox{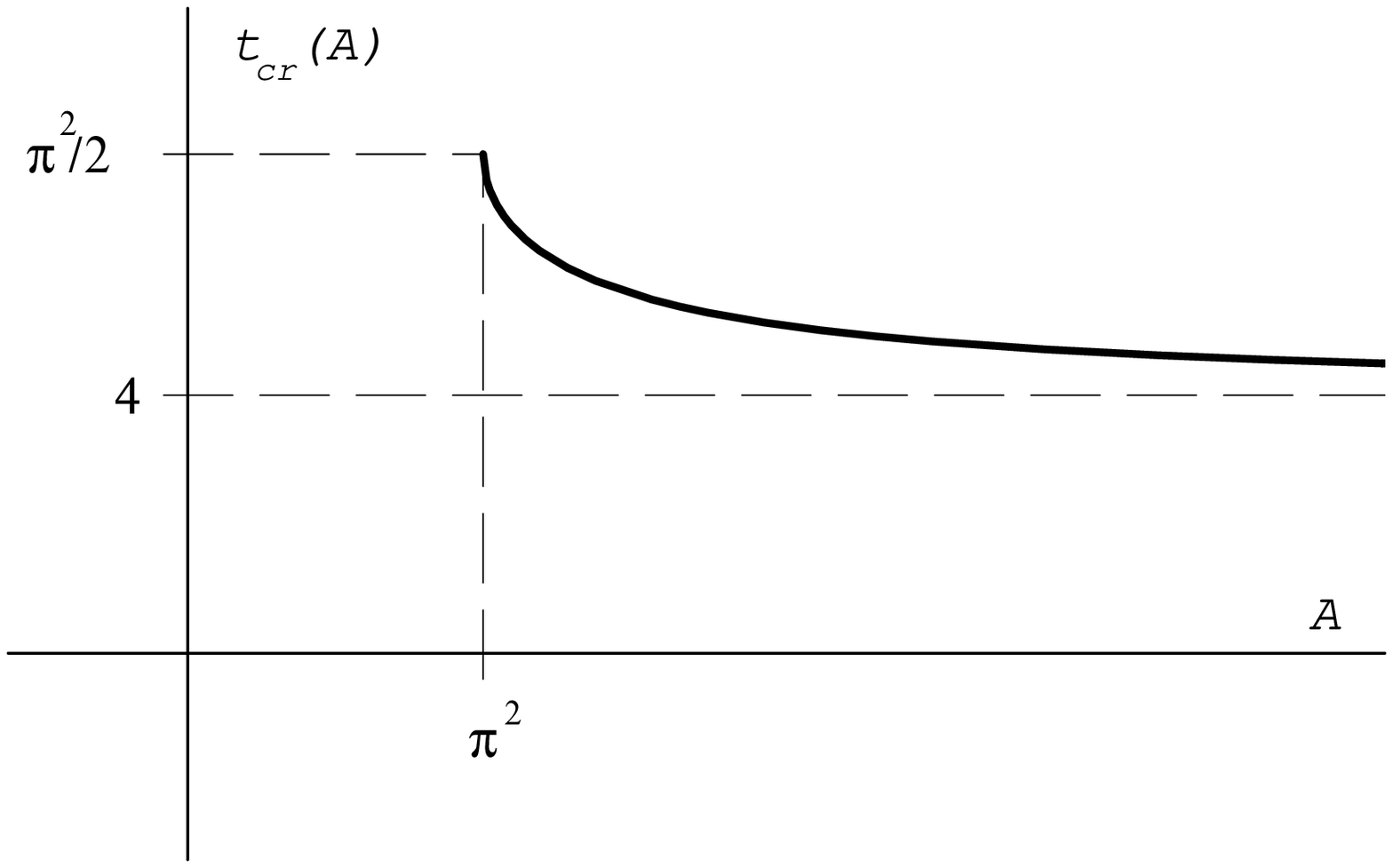}}

The dependence of the critical Wilson loop size $t_{\rm cr}=
{1\over 2}A-\tau_0(A)$ on $A$ is shown in \tcraplot.
Most remarkably, $t_{\rm cr}$ tends to a finite limit
equal to $t_{\rm cr}=4$ as $A$ goes to infinity.
Thus even in the QCD on a plane the eigenvalue distributions
for small contours have a gap which disappears as the area
enclosed by the contour passes through $4$ \REFdol. That is to say,
the master field for QCD on a plane contains some trace of
the Douglas--Kazakov phase transition and of the weak
coupling phase.

Since the asymptotic property $t_{\rm cr}\to 4$ will be important to
us later on, let us prove it.
Making the substitution $\lambda=b\cos \varphi$ in \tauzero\ we get
$$\tau_0(A)={2\over a}\int\limits_0^{\pi/2}{\cos^2\varphi\, d\varphi
\over \sqrt{\kappa^2+\sin^2\varphi}}$$
where $\kappa^2={a^2\over b^2}-1={{k^{\prime}}^2\over k^2}$.
At $A\to\infty$, as a consequence of \ellpar, ${{k^{\prime}}^2}\simeq
16\ee^{-A/4}\to 0$ and thus $\kappa^2=16\ee^{-A/4}+\dots$. In this
limit we can break up our integral into two parts and construct the asymptotic
expansions for both of them:
$$\eqalign{
&\int\limits_0^{\pi/2}{\cos^2\varphi\, d\varphi
\over \sqrt{\kappa^2+\sin^2\varphi}}=
\int\limits_0^{\epsilon}{d\varphi\over \sqrt{\kappa^2+\varphi^2}}+
\int\limits_{\epsilon}^{\pi/2}{\cos^2\varphi\over \sin\varphi}\, d\varphi
\cr&={\rm arsinh}{\epsilon\over \kappa}+\ln {\rm cotan}\,{\epsilon\over 2}
-1 +{\cal O}(\epsilon)+{\cal O}(\kappa^2)\cr
&=-1+\ln{4\over\kappa}+\dots=-1+{A\over 8}+ {\cal O}(\ee^{-A/4})\cr}$$
In these formulas we chose an auxiliary small quantity $\epsilon$ so that
$\kappa\ll\epsilon\ll 1$. Then at the end $\epsilon$ drops out
of the final result, leaving us with the desired asymptotics.

Now from \ellpar\ $a={1\over 2}+{\cal O}(\ee^{-A/4})$ and thus finally
$$\tau_0(A)={A\over 2}-4 +{\cal O}(\ee^{-A/4})$$
leading to
\eqn\tcritical{
t_{\rm cr}(A) =4+{\cal O}(\ee^{-A/4})}
as claimed.

Let us now demonstrate that the asymptotic behaviour of Wilson loops
for large winding numbers $n$ is related to the Douglas--Kazakov
transition. The most convenient way to do this is to continue
considering the planar case. For one, as we saw, the QCD on
a plane retains some trace of the weak coupling phase (the gap
in the eigenvalue distribution for loops with area less than 4).
On the other hand, all Wilson loops on a plane can be
evaluated explicitly, making the analysis very simple.

Indeed, the Wilson loops on a plane equal \REFkazkos\REFrossi
\eqn\wlplane{
W_n(A)=\Bigl\langle{1\over N} {\rm Tr}\, U_C^n\Bigr\rangle=
{1\over n}\ee^{-{nA\over 2}}L_{n-1}^{(1)}(An)}
where $L_{n-1}^{(1)}(x)$ are the so-called associated Laguerre
polynomials
\eqn\laguerrep{
L_{n-1}^{(1)}(x)=\sum_{k=0}^{n-1}(-x)^k{n!\over (n-k-1)! \,k!\,
(k+1)!}
=\oint {dt\over 2\pi i}\Bigl(1+{1\over t}\Bigr)^n \ee^{-x t}.}
Using this expression we can investigate the behaviour of $W_n(A)$
as $n\to\infty$ at fixed $A$. We will see that $W_n(A)$ decays
exponentially  in $n$ if $A>4$, but as a power of $n$   if $A<4$.
To prove this we need to know the asymptotic behaviour of the
Laguerre poynomials
as $x=An\to\infty$. This can be found by treating the integral in
\laguerrep\ using the saddle point method\foot{This
calculation is remarkably similar to the calculation of instanton
contributions in the weak coupling phase which are expressed as
Laguerre polynomials as well \REFinst.  This is not suprising since
these are different
manifestations of the same phase transition.}. The asymptotics of this
integral is governed by the minimum of the functional
\eqn\func{
\Phi(t)={x\over n}t -\ln \Bigl(1+{1\over t}\Bigr).}
This minimum is reached at
$$t_{\pm}=-{1\over 2}\pm \sqrt{{1\over 4}-{n\over x}}=
-{1\over 2}\pm \sqrt{{1\over 4}-{1\over A}}.$$
If $A>4$ then both of $t_+$ and $t_-$ are real. Inn this case it is possible
to show that the dominant contribution comes from the region around
$t=t_+$ and equals
\eqn\lagstr{
L_{n-1}^{(1)}(x)\simeq {1\over\sqrt{2\pi n \bigl|\Phi^{\prime\prime}(t_+)
\bigr|}}\ee^{-n \Phi(t_+)}=
{A\over \sqrt{2\pi n}}\Bigl(1-{4\over A}\Bigr)^{-{1\over 4}}
\ee^{{nA\over 2}[1-\gamma(4/A)]}}
where we  have introduced a special function
\eqn\gammax{
\gamma(x)=\sqrt{1-x}-{x\over2}\ln{1+\sqrt{1-x}\over 1-\sqrt{1-x}}=
2\sqrt{1-x}\sum_{s=1}^{\infty}{(1-x)^{2s}\over 4s^2-1}>0.}
Thus at $A>4$
\eqn\wllpinftystrong{
W_n(A)\simeq(-)^{n-1}{A\over \sqrt{2\pi n^3}}
\Bigl(1-{4\over A}\Bigr)^{-{1\over 4}}
\ee^{-{nA\over 2}\gamma\bigl({4\over A}\bigr)}}
so that as $n\to\infty$, $W_n(A)$ decays exponentially with the
index $\alpha(A)={A\over 2}\gamma({4/A})>0$.

On the contrary, if $A<4$ the saddle points $t_+$ and $t_-$ are
complex conjugate and make comparable contributions.
Then
\eqn\lagweak{\eqalign{
L_{n-1}^{(1)}(x)&\simeq(-)^{n-1}
{1\over \sqrt{2\pi n \bigl|\Phi^{\prime\prime}(t_+)
\bigr|}}2 {\rm Re}\, \Bigl\{\ee^{-n \Phi(t_+)-i \pi/4}\Bigr\}\cr
&=(-)^{n-1}\sqrt{{2\over \pi n}}\, A\, \Bigl({4\over A}-1
\Bigr)^{-{1\over 4}}
\ee^{nA\over 2}\cos\Bigl\{{nA\over 2} \Gamma\Bigl({4\over A}
\Bigr) +{\pi\over 4}\Bigr\}\cr}
}
where $\Gamma(x)=\sqrt{x-1}-x {\rm arctan}\sqrt{x-1}$.
As a result, for $A<4$ the Wilson loops
\eqn\wllpinftyweak{
W_n(A)\simeq\sqrt{{2\over \pi n^3}}A \Bigl({4\over A}-1\Bigr)^{-{1\over 4}}
\cos\Bigl\{{nA\over 2} \Gamma\Bigl({4\over A}
\Bigr) +{\pi\over 4}\Bigr\}
}
do not decay exponentially. Rather, they oscillate with $n$ and
the amplitude of oscillation decays as $n^{-3/2}$. Such behaviour
is similar to what we encounter in the weak coupling phase on a
sphere before the Douglas--Kazakov transition occurs\foot{Let us
emphasize that the absence of a gap and the exponential decay
of Wilson loops are not automatic consequences of each other. In fact,
the $n\to\infty$ behaviour of $W_n(A)$ probes the degree of smoothness
of the eigenvalue distribution. Since when there is no gap the distribution
is infinitely smooth the Wilson loops decay faster than any power of $n$.
On the other hand, if we do have a gap, then at the edge of the
support of the eigenvalues higher derivatives become discontinuous, and
the loops decay powerlike.}.

To summarize, the phase transition in continuum two-dimensional
QCD exibits itself through physical observables in several ways.
First, the density of eigenvalues for Wilson loop matrices develops
a gap in the weak coupling phase. Second, the asymptotic behaviour
of Wilson loops with large winding numbers is different in the
two phases. Finally, the Young tableau of the dominant representation
also can be expressed in terms of physical observables using the
duality formula \ssupsim.

\newsec{Instanton contributions to the Wilson loops.}

As we  saw in  the previous section, the  Douglas--Kazakov phase
transition is associated with the development of a gap in the
eigenvalue distributions for Wilson matrices.  In this section we
will argue that this phenomenon, very much like the original
Douglas--Kazakov transition on a sphere at $A_{\rm cr}=\pi^2$,
is induced by instantons.

Indeed, it is known that the  partition function of QCD$_{2}$
can be repesented exactly as a  sum of  contributions localized at
the instantons, that is, the classical solutions of the
theory \REFwitte\foot{By
instantons we mean all possible solutions  of the classical
Yang--Mills equations even if they are unstable.}. As it turns out,
in  the weak coupling phase, at $A_{\rm cr}<\pi^2$, the dominant term
comes from the expansion around the perturbative vacuum,
${\cal A}_{\mu}(x)=0$, while in the  strong coupling phase,
$A_{\rm cr}>\pi^2$, a certain
nontrivial instanton configuration will dominate.

A similar analysis can be carried out for Wilson loops. We can anticipate
that the average of a Wilson  loop will be given just by its classical
value as determined by  the dominant
instanton configuration\foot{This is the value we
obtain if we substitute the classical ${\cal A}_{\mu}(x)$ for the
dominant instanton into ${\rm tr}\, U= {\rm tr}\, {\cal P}{\rm exp}\oint
{\cal A}_\mu (x) \thinspace d x^\mu$.} modified  by quantum
corrections. The easiest way to determine these is to perform a
Poisson resummation in the original  formula  for Wilson loops
in terms of the representations of the gauge group. On the sphere,
for a simple contour which divides it into two patches of
areas $A_1$ and $A_2$, the Wilson loop is given by
\eqn\wlb{
\eqalign{
W_n(A_1, &A_2)\equiv\Bigl\langle{1\over N}{\rm tr}U^n\Bigr \rangle
={1\over {\cal Z}(A_1+A_2)}\sum_{R,S} d_R d_S \cr
&\times\biggl[{1\over N}\int\, dU
\, ({\rm tr}\, U^n)\, \chi_R(U)\chi_S(U^{\dagger})\biggr]
\ee^{-{A_1\over 2N}C_2(l^{(R)})-{A_2\over 2N} C_2(l^{(S)})}.\cr}}
In this formula $R$ and $S$ label all irreducible representations
of the gauge group, $d_R$ and $d_S$ are their dimensions and
$\chi(U)$ are the characters given by \char. \wlb\
can be derived easily by gluing two discs of areas
$A_1$ and $A_2$ along the loop.

Labelling $R$  and $S$ by two sets of integers, $l^{(R)}_1>l^{(R)}_2
>\dots>l^{(R)}_N$ and $l^{(S)}_1>l^{(S)}_2>\dots>l^{(S)}_N$ and
transforming $\int dU$ to an integral over the eigenvalues of $U$
we get
\eqn\wlres{
\eqalign{
W_n(A_1, A_2) &{\cal Z}(A_1+A_2)=\kern-0.7em
\sum_{\scriptstyle{l^{(R)}_1>l^{(R)}_2>\dots>l^{(R)}_N
\atop
l^{(S)}_1>l^{(S)}_2>\dots>l^{(S)}_N}}
\kern-1.5em\Delta(l_i^{(R)})\Delta(l_j^{(S)})
\ee^{-{A_1\over 2N}C_2(l^{(R)})-{A_2\over 2N} C_2(l^{(S)})}\cr
&\times \int d\theta_1\dots d\theta_N \, \biggl(
{1\over N} \sum_{k=1}^N \ee^{i n \theta_k}\biggr)\,
{\rm det} \bigl|\!\bigl|\, \ee^{i l_i^{(R)}\theta_p}\bigl|\!\bigl|\,
{\rm det} \bigl|\!\bigl|\, \ee^{-i l_j^{(S)}\theta_q}\bigl|\!\bigl|,
\cr}}
with $\Delta(x_i)\equiv \prod_{i<j}(x_i - x_j)$. Now we can expand the
determinants and use the antisymmetry of $\Delta(l)$ with respect to
the permutation of $l_i$'s to remove the ordering restriction on
the $l$'s. We obtain
\eqn\wzun{\eqalign{
W_n&(A_1, A_2) {\cal Z}(A_1+A_2)=
{1\over N}
\sum_{k=1}^N \sum_{l_i^{(R)}, l_j^{(S)}\atop
{\scriptstyle unrestricted}} \Delta(l_i^{(R)})\Delta(l_j^{(S)})\cr
&\times\ee^{-{A_1\over 2N}C_2(l^{(R)})-{A_2\over 2N} C_2(l^{(S)})}
\int d\theta_1\, d\theta_2\dots d\theta_N\,
\ee^{in \theta_k}\prod_{p=1}^N \ee^{i l_p^{(R)} \theta_p}
\prod_{q=1}^N\ee^{-il_q^{(S)}\theta_q}\cr
&={1\over N}\sum_{k=1}^N\sum_{l_i^{(R)}} \Delta(l_i^{(R)})\Delta(l_j^{(S)})
\, \ee^{-{A_1\over 2N}C_2(l^{(R)})-{A_2\over 2N} C_2(l^{(S)})}
}}
where now the summation is only over $l_i^{(R)}$, the $l_j^{(S)}$
being determined by $l_i^{(S)}= l_j^{(R)}+n \delta_{j, k}$.
That is to say, $l_i^{(S)}= l_j^{(R)}$ for all $i$ except for $i=k$, when
$l_k^{(S)}= l_k^{(R)}+n$.

To represent \wzun\ as a sum of instanton contributions let us
transform it using the Poisson resummation formula,
\eqn\poiss{
\sum_{n_1, \dots , n_N=-\infty}^{+\infty}f(n_1, \dots , n_N)=
\sum_{m_1, \dots , m_N=-\infty}^{+\infty}F(2\pi m_1, \dots , 2\pi m_N)}
where
$$F(p_1, \dots , p_N)=\int\limits_{-\infty}^{+\infty}f(x_1, \dots, x_N)
\, \ee^{i(p_1 x_1 +\dots +p_N x_N)} dx_1\dots dx_N.$$
First we recall that, up to an irrelevant uniform shift  of $l_k$'s
by $(N-1)/2$,
$$C_2(l)=\sum_{i=1}^N l_i^2.$$
Thus
$$\eqalign{
A_1 C_2(l^{(R)})&+A_2 C_2(l^{(S)})\cr
&= {A_1+A_2\over 2}\bigl[
C_2(l^{(R)})+C_2(l^{(S)})\bigr]+{A_1-A_2\over 2}\bigl[
C_2(l^{(R)})-C_2(l^{(S)})\bigr]\cr
&={A_1+A_2\over 2}\bigl[
C_2(l^{(R)})+C_2(l^{(S)})\bigr]+{A_2-A_1\over 2}(n^2+2nl_k^{(R)}).\cr}$$
and we can represent our Wilson loops in the form
$$\eqalign{
&W_n(A_1, A_2) {\cal Z}(A_1+A_2)\cr
&=\ee^{-{n^2\over 4N}(A_2-A_1)}{1\over N}
\sum_{k=1}^{N} \sum_{l_i^{(R)}}
\varphi\bigl(\{l_i^{(R)}\}\bigr)
\varphi\bigl(\{l_i^{(R)}+\delta_{i, k}n\}\bigr)\,
\ee^{-{n\over 2N}(A_2-A_1)l_k^{(R)}},\cr}$$
where we have introduced a new function of $N$ variables $\{l_j\}$
\eqn\vph{
\varphi\bigl(\{l_j\}\bigr)=\Delta(l_j)\, \ee^{-{A\over 4N}\sum_{p=1}^N
l_p^2},}
$A=A_1+A_2$ being the total area of the sphere.

To apply the Poisson formula \poiss\ we have to find the Fourier
transform of the function under summation:
$$\eqalign{
F&(p_1, \dots, p_N)\cr
=&\int\limits_{-\infty}^{+\infty}dx_1\dots dx_N\,
\varphi\bigl(\{x_i\}\bigr)\, \varphi\bigl(\{x_i+\delta_{i, k} n\}\bigr)
\, \ee^{-{n\over 2N}(A_2-A_1)x_k} \, \ee^{i(p_1x_1+\dots+p_Nx_N)}\cr
=&\int\limits_{-\infty}^{+\infty}dx_1\dots dx_N\,
\varphi\bigl(\{x_i\}\bigr)\, \varphi\bigl(\{x_i+\delta_{i, k} n\}\bigr)
\, \ee^{i({\tilde p}_1 x_1+\dots+{\tilde p}_N x_N)}\cr
}$$
where ${\tilde p}_j=p_j+ (in/2N) (A_2-A_1)\delta_{i, k}$.
Recalling that the Fourier transform of a product is a convolution of
the individual Fourier transforms we get
\eqn\FFt{
F(p_1, \dots, p_N)=\int\limits_{-\infty}^{+\infty}dy_1\dots dy_N \,
\psi\Bigl(\Bigl\{{{\tilde p}_j+y_j\over 2}\Bigr\}\Bigr)\, \psi
\Bigl(\Bigl\{{{\tilde p}_j-y_j\over 2}\Bigr\}\Bigr)
\, \ee^{-i({\tilde p}_k-y_k)n/2},}
where $\psi\bigl(\{p\}\bigr)$ is the Fourier transform of $\varphi\bigl(\{
x\}\bigr)$,
\eqn\ftrpsi{
\psi\bigl(\{p\}\bigr)\equiv\int\limits_{-\infty}^{+\infty}dx_1\dots dx_N\,
\varphi\bigl(\{x_i\}\bigr)\, \ee^{i(p_1 x_1+\dots +p_N x_N)}
=C_N \Delta(p_i) \, \ee^{-{N\over A}\sum_{i=1}^N p_i^2}}
with some $N$-dependent constant which will cancel  later on\foot{To
derive this one represents $\Delta(l_j)$ in \vph\ as a Van der Monde
determinant and performs the integrations explicitly. The answer can be
simplified to give \ftrpsi\ if we remember  that
one can add rows of a determinant without changing it. See \REFminahan\
for the details.}.
The phase factor in \FFt\ is due to the shift of $x_k\to x_k+n$ in
$\varphi\bigl(\{x_i+\delta_{i, k} n\}\bigr)$.
Thus
$$\eqalign{
F(p_1&, \dots, p_N)={\tilde C}_N\, \ee^{-i {n {\tilde p}_k/ 2}}\,
\ee^{-{N\over 2A}\sum_{j=1}^N p_j^2}\cr
&\int\limits_{-\infty}^{+\infty}dy_1\dots dy_N \,\biggl[\prod_{i<j}
({\tilde p}_{i j}^2-y_{i j}^2)\biggr]\, \ee^{i {n y_k/ 2}}\,
\ee^{-{N\over 2A}\sum_{j=1}^N y_j^2} \cr}$$
where ${\tilde p}_{ij}\equiv {\tilde p}_i-
{\tilde p}_j, \  y_{ij}\equiv y_i-y_j$ and
${\tilde C}_N$ is another constant, ${\tilde C}_N= C_N^2/2^{N(N-1)}$.

Combining the pieces we finally obtain the following representation
for the Wilson loops:
\eqn\irwl{\eqalign{
&W_n(A_1, A_2) {\cal Z}(A_1+A_2)\cr
&={1\over N}\sum_{k=1}^N \sum_{\scriptstyle m_1,
\dots, m_N=-\infty}^{\scriptstyle +\infty} F\Bigl(2\pi m_j+\delta_{j, k}{in
\over 2N}(A_2-A_1)\Bigr)\cr
&={1\over N}\sum_{k=1}^N \sum_{\scriptstyle m_1,
\dots, m_N=-\infty}^{\scriptstyle +\infty}\ee^{-{2\pi^2 N\over A}\sum_{j=1}^N
m_j^2}
\, \ee^{-2\pi i n m_k{A_2/ A}}\,\ee^{{n^2}(A_1-A_2)^2/8AN}\cr
&\times\Biggl\{\int\limits_{-\infty}^{+\infty}dy_1\dots dy_N \,
\ee^{in {y_k/ 2}}\prod_{i<j}\bigl[4\pi^2 {\tilde m}_{ij}^2-
y_{ij}^2\bigr]\, \ee^{-{N\over 2A}\sum_{j=1}^N y_j^2}\Biggr\}\cr}}
where ${\tilde m}_{ij}\equiv {\tilde m}_i-{\tilde m}_j$ and ${\tilde m}_j=
m_j+in\delta_{j, k}(A_2-A_1)/4\pi N$.

This formula has an interpretation in terms of instantons. Indeed,
the instantons of QCD$_2$ can be labelled by $N$ integers $m_1,
\dots, m_N$ and have the action
\eqn\iac{
S_{\rm inst}(m_1, \dots, m_N)= {2\pi^2 N\over A}\sum_{j=1}^N
m_j^2.}
The corresponding field configuration is just a collection of
Dirac monopoles
\eqn\instanton{
A_{\mu}(x)=\left(\matrix{m_1 {\cal A}_{\mu}^{0}(x) & 0 & \ldots & 0 \cr
                         0      & m_2 {\cal A}_{\mu}^{0}(x) & \ldots &0\cr
                         0 &\ldots&\ldots&0\cr
                         \vdots&\vdots&\ddots&\vdots\cr
                         0& 0 &\ldots &m_N{\cal A}_{\mu}^{0}(x)\cr
}\right)
}
where ${\cal A}_{\mu}^{0}(x)={\cal A}_{\mu}^{0}(\Theta, \phi)$ is the Dirac
monopole potential,
$${\cal A}_{\Theta}^{0}(\Theta, \phi)=0 , \quad\   {\cal A}_{\phi}^{0}
(\Theta, \phi)={1-\cos \Theta\over 2}, $$
$\Theta$ and $\phi $ being the polar (spherical) coordinates on $S^{2}$.
We can see that the terms $\ee^{2\pi i n m_k{A_2/ A}}$ in \irwl\ represent the
classical contribution of the field configurations \instanton\ to
the Wilson loop while the term
\eqn\zin{
\zeta_n^{(k)}(m_1, \dots, m_N)=\int\limits_{-\infty}^{+\infty}dy_1\dots dy_N \,
\ee^{in {y_k/ 2}}\prod_{i<j}\bigl[4\pi^2 {\tilde m}_{ij}^2-
y_{ij}^2\bigr]\, \ee^{-{N\over 2A}\sum_{j=1}^N y_j^2}}
represents quantum corrections due to the fluctuations around the
instanton.
Thus our final result can be represented in the form
\eqn\fire{
W_n(A_1, A_2)={1\over {\cal Z}(A)}\kern-0.2em
\sum_{\scriptstyle all\atop \scriptstyle
instantons}\kern-0.7em
\ee^{-S_{\rm inst}} \biggl[{1\over N}\sum_{k=1}^N
\ee^{-2\pi i n m_k{A_2/ A}}\zeta_n^{(k)}\biggr]\, \ee^{{n^2}(A_1-A_2)^2/8AN}.}

In fact, if we consider \irwl\ with $n=0$, we will reproduce
the instanton representation for ${\cal Z}(A)$ constructed by
Minahan and Polychronakos \REFminahan,
\eqn\witten{
{\cal Z}=\sum_{m_1,\ldots ,m_N=-\infty}^{+\infty}w(m_1, \ldots, m_N)
\thinspace\ee^{-S(m_1,\ldots ,m_N)}
}
where
\eqn\minahan{
w(m_1, \ldots, m_N)=\int_{-\infty}^{+\infty}\prod_{i=1}^{N}
\thinspace\prod_{i<j=1}^{N}
\big(4\pi^2 m_{ij}^2-y_{ij}^2\big)\thinspace
\ee^{-{N\over 2A}\sum_{i=1}^{N}y_i^2}
}
is the statistical weight of the configuration $(m_1, \dots, m_N)$.

In the large $N$ limit we expect that the sum in \witten\ is
dominated by a certain
configuration $(m^{(0)}_1, \dots, m^{(0)}_N)$ depending on the area $A$.
(For example, in the weak coupling  phase the  dominant configuration is
$m^{(0)}_1=\dots=m^{(0)}_N=0$.) As $N\to\infty$
for any  such configuration  the integrals
\zin\ and \minahan\ are dominated by some saddle points in $y_i$'s. Moreover,
the  saddle  points for these two integrals are exactly the same,
as they are determined by the balance of $\ee^{{\cal  O}(N^2)}$ terms
which are common  for both integrals. Thus, in the large $N$ limit,
we can determine $\zeta_n^{(k)}/w$ by substituting
the values  $y_i=y_i^{(0)}$  into the ratio of the integrands:
$${\zeta_n^{(k)}\over w}= \ee^{i n {y_k^{(0)}/ 2}}
\prod_{i<j}\biggl( 1-{4\pi^2(m_{ij}^2-{\tilde m}_{ij}^2)\over
4\pi^2m_{ij}^2- y_{ij}^{(0)2}}\biggr).$$
Note that, if $A_1=A_2=A/2$ (that  is, if we are considering
the Wilson  loop around the equator) then $m_i={\tilde m}_i$ and
$${\zeta_n^{(k)}\over w}=\ee^{i n {y_k^{(0)}/ 2}}.$$
Therefore in the leading large $N$ order
$$W_n(A_1, A_2)={1\over N}\sum_{k=1}^N \ee^{i\pi n m_k}{\zeta_n^{(k)}\over
w}={1\over N}\sum_{k=1}^N \ee^{i n (\pi m_k +y_k^{(0)}/2)}.$$
Thus we see that the large $N$ density $r(\eta)$ describing the
distribution of numbers $\eta_k=y_k^{(0)}/2+\pi m_k$ is nothing but
the eigenvalue distribution for the equatorial Wilson loop, $\sigma_0(\theta)$.
In the weak coupling phase, where all $m_k=0$, this identity is easy  to
check directly.

Let us also mention that the distribution of $(m^{(0)}_1, \dots, m^{(0)}_N)$
for the dominant instanton  configuration in  the strong coupling phase
can be determined from the formula
\eqn\dominst{
\sum_{m=-\infty}^{+\infty}p_m \ee^{-imq}=\int\limits_{-b}^{+b}dh\,
\exp\biggl\{-{q\over \pi}
\biggl[-\kern-1.1em\int {\rho_Y(y)\, dy\over h-y}-Ah\biggr]\biggr\}
=1-2b+\int\limits_{-b}^{+b}dh\,\ee^{iq\rho_Y(h)}}
where $q\in [-\pi, \pi]$,
$p_m$ is the probability to find an integer $m$ among the set
$\{m^{(0)}_1, \dots, m^{(0)}_N\}$ and $\rho_Y(y)$ is the Young tableau
density given by \strsph. This can be derived using the Poisson resummation
in a way similar to \irwl.

As we can see from \fire,
the quantum corrections make important quantitative
contributions to Wilson loops.
However, the  physics of the phase transition can be
understood even if we neglect them. Consider the classical
contribution to the Wilson loop,
\eqn\clwl{W^{\rm cl}(A_1, A_2)={1\over N}\sum_{k=1}^N \ee^{-2\pi
i n m_k A_2/A}.}
The effective eigenvalue distribution for such a loop will have a gap if the
angles $\theta_k=2\pi m_k A_2/A$ are restricted to a certain domain
inside of a circle, $|\theta_k|\le \theta_{\rm max}<\pi$ for all $k$.
Since $ \theta_{\rm max}= 2\pi m_{\rm max} A_2/A$, we get
the condition on the area of our loop
$$A_2<{A\over 2 m_{\rm max}}.$$
As $A_2$ increases this inequality may  be violated and the eigenvalue gap
will disappear\foot{On the other hand, in the  weak coupling phase all
$m_k=0$ and the inequality $A_2 m_{\rm max}<A/2$ is always satisfied,
consistent with the fact that there is always a gap.}.
The exact value of $A_2$ when this will occur will be
smaller than $A/2 m_{\rm max}$ since quantum fluctuations cause some
additional widening of the eigenvalue distribution.

In general, for  $A_2=0$ we  always have $\theta_k=0$, even in
four-dimensional QCD, since the  corresponding Wilson matrix is the
identity. On the  other hand, as the area of the  loop becomes very
large the gauge fields at the distant points  of the  contour are
uncorrelated and the distribution  of angles becomes uniform:
$\sigma(\theta)=1/2\pi.$ This is also true in the four-dimensional
case. We can
see that while zero-area loops have an eigenvalue gap, the infinitely
large  ones do not. This might be a piece of  evidence for a large $N$
phase transition in QCD in $any$ dimension. However, we also saw that
the quantum corrections tend to broaden the eigenvalue distribution.
In fact, it might happen that for  any small  nonzero area the
distributions develop infinitesimal tails and do not have a  gap
anymore. The nontrivial fact about QCD$_2$ is that such a phenomenon  does
not occur there. The reason is that the Hopf equation, governing the
evolution  of  the eigenvalue densities, essentially prohibits
the  development of such smooth tails. Rather, its  generic solution
has a (moving) edge and vanishes as $\sigma(\theta, t)\sim\sqrt{\theta-
\theta_{cr}(t)}$ near it. Whether or not the situation  is  similar in
higher dimensions is unclear.

One could contemplate analysing the 4d Wilson loops for small area
by perturbation theory, since in this case the asymptotic freedom
applies. However, to obtain the eigenvalue distribution one would
have to evaluate the Wilson loops with arbitrarily high winding numbers
$n$. This corresponds to multiplying the charge flowing around the loop
by $n$, and for high $n$ this takes us out of the domain  where we can
rely on perturbation theory.

Let us emphasize that the phase transition in the eigenvalue densities
of the  Wilson loops does  not yet imply that there should be a
singularity in the free energy of the theory. For example, this is the case
with QCD on a two-dimensional plane. But such a  phenomenon might imply
a phase transition for the free energy of QCD in a  box of a finite
size\foot{See also the recent paper by Douglas \REFdgtr\ who draws similar
conclusions.}
(say, for a sphere in the case of QCD$_2$). In general, such transitions would
mean that the string representation of QCD$_4$, if constructed, will have
only a certain domain of validity. On the other hand, it may offer an
explanation how the short distance behavoiur of QCD which  is described
in terms of particles, agrees with the string description.

\newsec{Exact critical area for the phase transition on a disc.}

As another application of the duality formula \ssupsim\ we will
calculate the critical area for   QCD on a disc. A disc can be
regarded as a particular case of a cylinder with one of the
boundary holonomy matrices set to identity, $U_{C_2}=I$, that is to
say $\sigma_2(\theta)=\delta(\theta)$.

Quite remarkably, in this case the boundary value  problem for the
Euler equations \bc, \euler\ can be reduced to a  Cauchy problem
and then solved explicitly using \hopfsol. To this end we need to
know how to specify the velocity $v(t=0,\, \theta)\equiv v_1(\theta)$
at the beginning of the trajectory so that at the end, when $t=A$,
the second boundary condition \bc\ is satisfied, $\sigma(\theta)|_{t=A}=
\sigma_2(\theta)=\delta(\theta)$.

While such a problem cannot be solved explicitly for an arbitrary
$\sigma_2(\theta)$, in the particular case $\sigma_2(\theta)
=\delta(\theta)$ the necessary velocity is given by
\eqn\exvel{
v_1(\theta)= -{\theta\over A} + -\kern-1.1em\int{\sigma_1(\theta^{\prime})
\, d\theta^{\prime}\over \theta-\theta^{\prime}}.}
To see why this is true let us recall that the boundary value  problem
identical to \bc, \euler\ occurs in the calculation of the
Itzykson--Zuber integral. The correspondence between these two
problems is established by $\sigma_1(\theta)\leftrightarrow\alpha(\theta),
\; \sigma_2(\theta)\leftrightarrow\beta(\theta)$ (see \pairf\ and
the analysis in the Appendix B). In the case $\sigma_2(\theta)
=\delta(\theta)$ we would have
$\beta(b)=\delta(b)$ which means that the argument $B$ of the
Itzykson--Zuber integral is simply  zero\foot{Note the important distinction
from the matrix $U_{C_2}$, which is the  $identity$  matrix for
$\sigma_2(\theta)=\delta(\theta)$.}. Then the integral itself is
obviously equal to one, $I_N(A, B=0)=1$. On the other hand, we can
apply the asymptotic formula \izas\ to obtain
\eqn\prinhamt{
S\bigl[\alpha(a), \beta(b)\equiv\delta(b)\bigr]=
-{1\over 2}\int \alpha(a)\, a^2\, da +{1\over 2}\int
\alpha(a)\,\alpha(a^{\prime})
\, \ln|a-a^{\prime}|\, da\, da^{\prime}.}
The functional $S$ is the action along the classical trajectory
connecting $\alpha(a)$ and $\beta(b)$ during the unit time interval,
$t=1$. The velocity at the beginning of the trajectory can be
found as
$$v(a)={\partial\over\partial a}{\delta S[\alpha, \beta]\over
\delta \alpha(a)}=-a + -\kern-1.1em\int{\alpha(a^{\prime})\, da^{\prime}
\over a-a^{\prime}}.$$
If the time interval is not equal to $t=1$, this formula can be easily
generalized\foot{To do this one considers the integral
$I_N(A, B|t)=\int dU  \, \ee^{{N\over t}{\rm Tr}(AUBU^{\dagger})}$.
Its asymptotics is obtained by replacing
${1\over 2}\int \alpha(a)\, a^2\, da\to
{1\over 2t}\int \alpha(a)\, a^2\, da$ and
${1\over 2}\int\beta(b) \, b^2\, db\to
{1\over 2t}\int\beta(b) \, b^2\, db$ in the formula \izas.}
$$v(a)=-{a\over t}+ -\kern-1.1em\int{\alpha(a^{\prime})\, da^{\prime}
\over a-a^{\prime}}$$
finally proving \exvel.

Since the Euler equations \euler\ are equivalent to the Hopf equation
\hopf, we can now set up the Cauchy problem for \hopf\ with the
initial condition
\eqn\initcon{
f(t=0, \theta)\equiv f_0(\theta)= -{\theta\over A}+
-\kern-1.1em\int{\sigma_1(\theta^{\prime})\, d\theta^{\prime}\over
\theta-\theta^{\prime}}+i\pi\sigma_1(\theta)=
-{\theta\over A}+\int{\sigma_1(\theta^{\prime})\, d\theta^{\prime}\over
\theta-\theta^{\prime}}.
}
Then the solution is determined by \hopfsol\ which for this particular
case translates into
\eqn\impeq{
\Bigl(1-{t\over A}\Bigr)f(t, \theta)=
-{\theta\over A}+\int{\sigma_1(\theta^{\prime})\, d\theta^{\prime}\over
\theta-\theta^{\prime}-tf(t, \theta)}.}
This equation can be used to determine $f(t, \theta)$ and, consequently,
$\sigma(t, \theta)$, for any specific $\sigma_1$. However, the problem
of the Douglas--Kazakov phase transition can be solved without doing so.
Indeed, as we proved in the previous subsection, in order to test for
the transition it is sufficient to find out whether or not the support
of $\sigma(t, \theta)$ develops a gap at any time $t\in[0, A]$. This
allows one to obtain an explicit formula for the area of the disk when the
transition takes place.

To simplify the analysis, let us consider the case of a symmetric
$\sigma_1(\theta)$, so that $\sigma_1(\theta)=\sigma_1(-\theta)$.
If the support of $\sigma_1$ is the whole circle $S^1=[-\pi, \pi]$
then in the process of evolution this support must always develop a gap.
Indeed, the final result of such evolution is a delta function with
support which covers only a single point $\theta=0$.

Such picture is characteristic of the strong coupling phase. Therefore,
in this situation we never have any weak coupling phase and the
transition is impossible. A different picture arises if the support
of $\sigma_1$ covers only a part of a circle $[-b, b]$ with $b<\pi$.
Then the Euler evolution can cause this support to expand with $t$ before
it starts contracting to $\theta=0$. If this expansion stops before
the edge of the support, $b(t)$, reaches $\pi$ then the system stays
in the weak coupling phase. Otherwise the transition to the strong coupling
phase will occur.

At the moment $t_c$ when $b(t)$ reaches its maximum $f(t_c,\, \theta=
b(t_c))=0$. Indeed, $\sigma(t_c, \theta)|_{\theta=b(t_c)}=0$ because
$b(t_c)$ is the endpoint of the support of $\sigma$. Moreover,
$v(t_c, b(t_c))=0$ because exactly at $t=t_c$ the support has just stopped
expanding and the velocity at its edge is zero\foot{Still, at $t=t_c$
the velocities $inside$ of the support do not have to vanish.}.
Thus we may determine $\theta_c=b(t_c)$ using \impeq\ with $f=0$:
$${\theta_c\over A}= \int{\sigma_1(\theta^{\prime})\, d\theta^{\prime}\over
\theta_c-\theta^{\prime}}.$$
The critical area $A=A_{\rm cr}$ corresponds to $\theta_c=\pi$ giving
\eqn\exar{
A_{\rm cr}={\pi\Biggl[ \int {\sigma_1(\theta)\, d\theta\over \pi-\theta}
\Biggr]^{-1}}.}
Note that the integral in \exar\ is not a
principal value integral. Rather, it is an ordinary integral
which has no singularities because $\theta=\pi$ lies outside of
the support of $\sigma_1$.

Like the duality relation formula \exar\ can be checked in  a number of
exactly solvable cases.For example,   QCD on a sphere corresponds to
$\sigma_1(\theta)=\delta(\theta)$ so that
$\int{\sigma_1(\theta)\, d\theta\over \pi-\theta}={1\over \pi}$.
Thus \exar\ yields $A_{\rm cr}=\pi^2$ reproducing the original
result of Douglas and Kazakov.

The same technique can be used to obtain the transition area
for a cylinder whenever we can evaluate the large $N$ limit of
the corresponding Itzykson--Zuber integral \izas\ with the eigenvalue
distributions $\alpha=\sigma_1$ and $\beta=\sigma_2$. This way we can
generalize \exar\ to the case of a $flat$ $\sigma_2$, that is,
$$\sigma_2(\theta)=\left\{
\eqalign{
&{1\over 2c_2}, \quad |\theta|\le c_2\cr
&0 \qquad {\rm otherwise}\cr}\right.$$
Then the $c_2\to 0$ limit would correspond to $\sigma_2(\theta)\to
\delta(\theta)$.

For such $\sigma_2$ the eigenvalues $\theta_j^{(2)}$ are
distrubuted uniformly over the interval $[-c_2, c_2]$:
$$\theta_j^{(2)}={2c_2\over N}\Bigl(j-{N\over 2}\Bigr)\qquad j=1\dots N$$
and the determinant in the Itzykson--Zuber formula \itzyk\
reduces to a Van der Monde determinant\foot{
We keep in mind the correspondence $\theta_k^{(1)}\leftrightarrow
a_k, \, \theta_j^{(2)}\leftrightarrow b_j$ and
$\alpha(\theta)\leftrightarrow \sigma_1(\theta),\,
\beta(\theta)\leftrightarrow\sigma_2(\theta)$.}
$$\eqalign{
&{\rm det}\Bigl|\!\Bigl|\ee^{{2c_2} a_k (j-{N\over 2})}
\Bigr|\!\Bigr|=
\ee^{-N c_2\sum_{k=1}^N a_k}{\rm det}\Bigl|\!\Bigl|\bigl(\ee^{{2c_2}
a_k}\bigr)^j\Bigr|\!\Bigr|=
\ee^{-N c_2\sum_{k=1}^N a_k}\, \Delta \bigl(\ee^{{2c_2}
a_k}\bigr)\cr
&=\exp\biggl[-N c_2\sum_{k=1}^N a_k +{1\over 2}\sum_{j\ne k=1}^N
\ln\Bigl|\ee^{{2c_2}a_j}-\ee^{{2c_2}a_k}\Bigr|\biggr].\cr}$$
Thus the large $N$ limit is easy to evaluate with the result that
the action functional $S$ equals\foot{See footnote 24 for the explanation
of $t$-dependence.}
\eqn\acfl{\eqalign{
S=&-{1\over 2t}\int \sigma_1(\theta)\, \theta^2 \, d\theta-
{c_2\over t}\int \sigma_1(\theta)\, \theta\, d\theta\cr
&+{1\over 2}\int\int \sigma_1(\theta)\, \sigma_1(\theta^{\prime})\,
\ln \bigl|\ee^{{2c_2\over t}\theta}-\ee^{{2c_2\over t}\theta^{\prime}}
\bigr|\, d\theta\, d\theta^{\prime}.\cr}
}
Therefore the  initial Hopf velocity solving the boundary problem \bc\
can be written down as
\eqn\invel{\eqalign{
&v_1(\theta)={\partial\over\partial \theta}{\delta S\over
\delta\sigma_1(\theta)}=-{\theta+c_2\over t}+{2c_2\over t}
-\kern-1.1em\int{\sigma_1(\theta^{\prime})\, d\theta^{\prime}\over
1-\ee^{{2c_2\over t}(\theta^{\prime}-\theta)}}\cr
&=-{\theta\over t}+{c_2\over t}-\kern-1.1em\int{
\sigma_1(\theta^{\prime})\, d\theta^{\prime}\over
{\rm tanh}\bigl[{c_2\over t}(\theta-\theta^{\prime})\bigr]}\cr}}
so that the initial function $f_0$ in \hopfsol\ is
$$f_0(\theta)=-{\theta\over t}+{c_2\over t}\int{
\sigma_1(\theta^{\prime})\, d\theta^{\prime}\over
{\rm tanh}\bigl[{c_2\over t}(\theta-\theta^{\prime})\bigr]}.$$

As before, the critical area $A_{\rm cr}$ can be found from
\eqn\creq{
f_0(\theta_c=\pi)\bigl|_{t=A_{\rm cr}}=0}
giving finally the formula
\eqn\crflat{
\int{
\sigma_1(\theta)\, d\theta\over
{\rm tanh}\bigl[{c_2\over A_{\rm cr}}(\pi-\theta)\bigr]}=
{\pi\over c_2}.}
In the limit $c\to 0$ this equation reproduces \exar.
On the other hand if $\sigma_1$ also is a flat distribution,
$\sigma_1(\theta)={1\over 2 c_1}$ for $|\theta|\le c_1$ then
\crflat\ gives us
\eqn\aflat{
{\rm tanh}{\pi c_1\over A_{\rm cr}}\, {\rm tanh}{\pi c_2\over A_{\rm cr}}=
{\rm tanh}{c_1 c_2\over A_{\rm cr}}.}

\newsec{Two-dimensional QCD on a vertex manifold.}

Another nontrivial problem  is the large $N$ QCD partition
function  on a vertex
manifold (\pantsfig), or pants diagram.  In a sense, the
cylinder and the ``vertex" are the only nontrivial
manifolds we have to consider. Any other two-dimensional surface
can be constructed by gluing together an appropriate number of
vertices and cylinders. Then, according to \convolu-\willoop\
the partition function on such composite manifold is just the
product of partition functions for its constituents. Moreover, we
can restrict ourselves to the limit  when the area of the vertex
manifold goes to zero. Indeed, we can create a vertex of any finite
area just by attaching cylinders to a vertex of infinitely small area.
In this case the partition function of QCD on a vertex becomes a
certain functional of the three eigenvalue distributions for the
boundary matrices\foot{Throughout this chapter these matrices will be
denoted simply as  $U_1$, $U_2$ and $U_3$.} $U_{C_1}$, $ U_{C_2}$ and
$U_{C_3}$.

This functional should be somewhat similar to a delta-function.
For one, in the case of a cylinder if $A\to 0$
\eqn\delcyl{
{\cal Z}_N(U_1, U_2 |A)\to \sum_{R}\chi_R(U_1)\chi_R(U_2^{\dagger})=
\delta_{\rm cl}(U_1, U_2)}
where $\delta_{\rm cl}$ is the delta-function on the set of conjugacy
classes of the  gauge group. That is to say, $\delta_{\rm cl}$ equates
the sets of eigenvalues of $U_1$ and $U_2$.
We can represent it in terms of the usual delta-function on a
group manifold $\delta(U)$:
\eqn\dcl{
\delta_{\rm cl}(U_1, U_2)=\int dU\, \delta(U_1UU_2^{\dagger}U^{\dagger}, I)}
with $I$ the identity matrix.

It is useful to think  of this function
as analogous to  the vertices of conventional field theory.  There, with every
vertex  there is  associated a delta-function of the total incoming momentum.
Obviously, something similar should appear in our case as well.
We expect that the structure of the partition function of
QCD on the vertex manifold will have two
ingredients. One is a selection condition (similar to the condition
that the sum of all incoming momenta is zero)
and the other ingredient  is the actual
value of the partition function when this condition is satisfied
(the  counterpart of the vertex coefficient in field theory).

To find the selection condition we start with the exact expression
for the partition function of ${\rm QCD}_2$ on the vertex manifold
of a finite area $A$
\eqn\fiar{
{\cal Z}_N(U_1, U_2, U_3|A)=\sum_R{\chi_R(U_1)\chi_R(U_2)\chi_R(U_3)\over
d_R}\ee^{-{A\over 2N}C_2(R)}.}
As $A\to 0$,
$${\cal Z}_N\to \sum_R{\chi_R(U_1)\chi_R(U_2)\chi_R(U_3)\over
d_R}\equiv {\cal Z}_N(U_1, U_2, U_3).$$
Using the formulas
\eqn\formchar{
\eqalign{
&\int dV\, \chi_R(AVBV^{\dagger})={1\over d_R}\chi_R(A)\chi_R(B)\cr
&\sum_R\chi_R(U)\chi_R(V^{\dagger})=\delta_{\rm cl}(U, V)\cr}}
we can rewrite this in the form analogous to \delcyl:
\eqn\delpants{
\eqalign{
{\cal Z}_N(U_1, U_2, U_3)&=\sum_R\int dV_1\, \chi_R(V_1 U_1 V_1^{\dagger}
U_2)\chi_R(U_3)\cr
&=\int dV_1\, dV_2\, \delta(V_1U_1V_1^{\dagger}V_2U_2V_2^{\dagger}U_3, I).\cr
}}
Since the characters in \fiar\ are class functions on the group
we can assume that $U_1$, $U_2$ and $U_3$ are diagonal matrices.

To evaluate the integral in \delpants\ we need to find  matrices
$V_1$ and $V_2$  that satisfy the equation
$V_1U_1V_1^{\dagger}V_2U_2V_2^{\dagger}U_3
=I$ and then compute the Jacobians necessary to integrate out
the delta-function.

First, we rewrite \delpants\ as
\eqn\delp{\eqalign{
{\cal Z}_N(U_1, U_2, U_3)=&\int dV_1\, dV_2\, \delta\bigl(V_1\bigl(U_1
(V_1^{\dagger}V_2)U_2(V_1^{\dagger}V_2)^{\dagger}\bigr)V_1^{\dagger},
U_3^{\dagger}\bigr)\cr
&=\int dV_1\, dM \delta\bigl(V_1(U_1MU_2M^{\dagger})V_1^{\dagger},
U_3^{\dagger}\bigr).\cr}}
Now, we can perform the integration with respect to $V_1$, keeping
$M$ fixed. The delta-function will pick up the specific value of $V_1=
V_1^{(0)}$ such that $V_1^{(0)}$ diagonalizes  the matrix
$Q(M)\equiv U_1MU_2M^{\dagger}$. Thus, after the integration over  $V_1$
the delta-function in \delp\ reduces  to a  delta-function equating
the eigenvalues of $Q(M)$ and $U_3^{\dagger}$. Indeed, if $V_1^{(0)}Q(M)
V_1^{(0)\dagger}=\Lambda$ where $\Lambda$ is diagonal then in the small
vicinity of $V_1^{(0)}$ we have $V_1=(I+v_1)V_1^{(0)}$ with some
antihermitian $v_1$ and
$$\eqalign{
&\delta(V_1Q(M)V_1^{\dagger}, U_3^{\dagger})=
\delta\bigl(\Lambda+[v_1, \Lambda], U_3^{\dagger}\bigr)\cr
&=\prod_{p<q}\delta\bigl({\rm Re}\, [v_1, \Lambda]_{pq}\bigr)
\delta\bigl({\rm Im}\, [v_1, \Lambda]_{pq}\bigr)
\, \prod_{p=1}^N \delta\bigl(\Lambda_{p}-(U_3^{\dagger})_p\bigr)\cr
&=\prod_{p<q}{\delta\bigl({\rm Re}\,(v_1)_{pq}\bigr)
\delta\bigl({\rm Im}\,(v_1)_{pq}\bigr)\over |\Lambda_{p}-
\Lambda_{q}|^2}\,
\prod_{p=1}^N \delta\bigl(\Lambda_{p}-(U_3^{\dagger})_p\bigr)\cr}$$
where $\Lambda_{p}$ and $(U_3)_p$ are the eigenvalues of the
(diagonal) matrices $\Lambda$ and $U_3$, respectively. Thus,
up to a $\Lambda$-dependent coefficient $c(\Lambda)$,
\eqn\zvoldm{
{\cal Z}_N(U_1, U_2, U_3)\sim\int dM \, c\bigl(\Lambda(M)\bigr)
\prod_{p=1}^N\delta\bigl(\Lambda_{p}(M)-(U_3)_p^*\bigr)}
where $\Lambda_{p}(M)$ are the eigenvalues of $Q(M)\equiv
U_1MU_2M^{\dagger}$. To estimate this integral we should
find $M=M_0$ such that $Q(M_0)$ has the same eigenvalues as
$U_3^{\dagger}$ and compute the Jacobian $J(M)={\rm det}\bigl|\!\bigl|
{\partial \Lambda_p(M)\over \partial M_{ab}}\bigr|\!\bigr|_{a, b=1}^N$.
The partition function ${\cal Z}_N$ will be inversely proportional
to this Jacobian.

If this Jacobian vanishes the partition function blows up.
Intuitively, in this case the volume of integration over $dM$
which contributes to \zvoldm\ is much larger than in the generic
case. The most singular situation is when not only the Jacobian
$J(M)$ bot the whole matrix ${\partial \Lambda_p(M)\over \partial M_{ab}}$
vanishes. Then any small shift away from $M=M_0$ does not violate the
condition imposed by the delta-function in \zvoldm\ and the
integration volume will be the largest. The condition on
$U_1$, $U_2$ and $U_3$ under which this occurs will be
an analogue of the condition $\sum_{i=1}^n
p_i=0$ imposed by the vertices in field theory.

To find this condition we must calculate the variation of eigenvalues
$\delta\Lambda_p(M)=\Lambda_p(M+\delta M)-\Lambda_p(M)$ to the first
order in $\delta M$ and require that such variation vanishes.

But $\Lambda_p(M)$ are the eigenvalues of the matrix $Q(M)$ with the
variation
$$\delta Q(M)=U_1\delta M U_2 M_0^{\dagger}-
U_1M_0 U_2 M_0^{\dagger} \delta M M_0^{\dagger}.$$
If
\eqn\diad{
V_1^{(0)} Q(M_0)V_1^{(0)\dagger}=
V_1^{(0)}U_1 M_0 U_2 M_0^{\dagger}V_1^{(0)\dagger}=\Lambda(M_0)}
is diagonal, then the variations of the diagonalizing matrices
$V_1^{(0)}$ do not contribute to the change in the eigenvalues
and, according to   first order perturbation theory
\eqn\cmt{\eqalign{
\delta \Lambda_p(M)&=\bigl(V_1^{(0)}\delta Q(M)V_1^{(0)\dagger}\bigr)_{pp}\cr
&=(V_1^{(0)}U_1\delta M U_2 M_0^{\dagger} V_1^{(0)\dagger})_{pp}-
(V_1^{(0)}U_1M_0U_2M_0^{\dagger}\delta M
M_0^{\dagger} V_1^{(0)\dagger})_{pp}\cr
&=\bigl(V_1^{(0)}U_1[\delta M M_0^{\dagger}, U_1M_0U_2M_0^{\dagger}]
V_1^{(0)\dagger}\bigr)_{pp}.\cr}}
If we introduce the matrices $\mu=\delta M M_0^{\dagger}$,
$\nu=V_1^{(0)}\mu V_1^{(0)\dagger}$ and $K_1=V_1^{(0)}U_1V_1^{(0)\dagger}$
and take into account that, due to \diad\ ,
$$M_0 U_2 M_0^{\dagger}=U_1^{\dagger}V_1^{(0)\dagger}\Lambda(M_0)
V_1^{(0)}$$
we obtain the following representation for \cmt:
\eqn\recps{
\delta \Lambda_p(M)=\bigl(K_1 \nu K_1^{\dagger}\Lambda(M_0)\bigr)_{pp}
-\bigl(\Lambda(M_0)\nu\bigr)_{pp}=
\bigl((K_1\nu K_1^{\dagger})_{pp}-\nu_{pp}\bigr)\Lambda_p(M_0)}
since $\Lambda(M_0)$ is a diagonal matrix.

The condition $\delta \Lambda_p(M)=0$ means that
$(K_1\nu K_1^{\dagger})_{pp}=\nu_{pp}$ for any $p$ and any antihermitian
$\nu=-\nu^{\dagger}$. This is possible only if $K_1$ is diagonal,
that is $(K_1)_{pr}=c_p\delta_{pr}$. Thus $V_1^{(0)}U_1V_1^{(0)\dagger}=
K_1$ where both $U_1$ and $K_1$ are diagonal matrices. This means that
$V_1^{(0)}$ is a permutation matrix and $c_p$ are the same as the
eigenvalues $(U_1)_p$ (maybe in a different order). In turn, this implies
the diagonality of $U_1^{\dagger}V_1^{(0)\dagger}\Lambda(M_0)V_1^{(0)}$
and, because of \diad, also the diagonality of $M_0U_2M_0^{\dagger}$.
Since $U_2$ is diagonal, $M_0$ also is a permutation matrix, and so is
$V_2^{(0)}=V_1^{(0)}M_0$.  In short, $V_1^{(0)}$, $V_2^{(0)}$ and
$M_0$ are some permutation matrices.

Going back to \delpants\ we conclude that $(V_1^{(0)}U_1V_1^{(0)\dagger}
)(V_2^{(0)}U_2 V_2^{(0)\dagger})U_3=I$. In other words, {\it when
properly ordered the eigenvalues of $U_1$, $U_2$ and $U_3$ must give unity
upon multiplication.}

This conclusion can be easily reformulated in terms of eigenvalue
distributions $\sigma_1(\theta)$, $\sigma_2(\theta)$ and $\sigma_3(\theta)$.
To do this, we introduce the ``numbering functions"
$\Sigma_j(\theta)$, $j=1, 2, 3$, such that $d\Sigma_j(\theta)/d\theta=
\sigma_j(\theta)$. These functions map the eigenvalue $\theta_i^{(j)}$
with number $i\in\{1, \dots, N\}$ to the fraction ${i\over N}\in[0, 1]$.
If we now construct the inverse mappings $V_j(u)$ so that
$\Sigma_j\bigl(V_j(u)\bigr)=u$ (no summation over $j$) then
the multiplication condition for eigenvalues\foot{Here we implicitly
presume that the eigenvalues have already been proprely ordered,
so that in fact $U_1 U_2 U_3=I$ is satisfied.}  $\lambda_i^{(1)}
\lambda_i^{(2)}
\lambda_i^{(3)}=1$
is equivalent to
\eqn\cdtn{V_1(u)+V_2(u)+V_3(u)=0.}
This is the selection condition imposed by the zero-area vertex  for large $N$.

This discussion  shows that the functions $V_j(u)$ are somewhat analogous to
 momenta in a conventional field theory. Moreover, using the formalism
of subsection 2.3 it is possible to show that the velocities $v_j(\theta)$
in the collective field representation satisfy a similar constraint,
\eqn\convel{
v_1\bigl(V_1(u)\bigr)=v_2\bigl(V_2(u)\bigr)=v_3\bigl(V_3(u)\bigr).}
This constraint together with
\def\s#1{\Sigma_#1\bigl(V_#1(u)\bigr)}
$$\s1=\s2=\s3$$
has a suggestive form. Let us recall that the purpose of the zero-area
vertices is to glue three cylinders together. The functions
$V_j$ provide mappings between the boundaries of these cylinders
which are similar to the mappings between different coordinate
patches forming the atlas of a manifold. This may mean that
the string theory of two-dimensional QCD is most naturally
formulated in terms of functions, functionally inverse to the
collective field variables $v$ and $\Sigma$.

Another application of \cdtn\ concerns the problem of the master field.
The collective field theory gives us a way of evaluating
the Wilson loops only without self-intersections. However, by joining
a number of nonintersecting loops we can obtain an arbitrarily complex
Wilson loop. Such joining is described by a vertex function we are
discussing. However, to evaluate the composite Wilson loop one
would need to know not merely the selection condition \cdtn,
but also the vertex function itself when this condition is
satisfied\foot{This would be a counterpart of the vertex coefficient
in field theory.}. This requires more subtle methods of calculation
and is a separate subject that will be discussed elsewhere.

\newsec{Acknowledgements.}

We are indebted to M. Douglas and V. Kazakov for very helpful conversations.

\appendix{A}{The change of variables in the collective
field description of ${\rm QCD}_2$.}

Our goal is to find the large $N$ limit of equation \predas.
To this end we set ${\tilde {\cal Z}}_N=\exp N^2{\tilde F}_N$
and transform \predas\ into an equation for ${\tilde F}_N$
with the result
\eqn\eqfn{\eqalign{
2{\partial {\tilde F}_N\over\partial A}= &{1\over N}
\sum_{k=1}^N {\partial^2 {\tilde F}_N\over \partial \theta_k^{(1)2}}
+{1\over N} \sum_{k=1}^N \biggl(N {\partial {\tilde F}_N\over
\partial \theta_k^{(1)}}\biggr)^2\cr&+
{2\over N} \sum_{k=1}^N U_k \biggl(N {\partial {\tilde F}_N\over
\partial \theta_k^{(1)}}\biggr)
+{1\over N^3} \sum_{k=1}^N {1\over {\cal D}(\theta^{(1)})}
{\partial^2\over
\partial \theta_k^{(1)2}} {\cal D}(\theta^{(1)})\cr}}
where
\eqn\cottang{
U_k\equiv {1\over N}{\partial\over \partial \theta_k^{(1)}}
\ln {\cal D}(\theta^{(1)})=
{1\over 2N} \sum_{j\ne k}{\rm cotan}{\theta_k^{(1)}-
\theta_j^{(1)}\over 2}.}

In the large $N$ limit all partial derivatives can be replaced
by the derivatives with respect to the eigenvalue densities,
$$N{\partial {\tilde F}_N\over\partial \theta_k^{(1)}}=
{\partial\over \partial\theta}{\delta {\tilde F}\over \delta
\sigma_1(\theta)}\biggl|_{\theta=\theta_k}$$
and all sums --- by the integrals
$${1\over N}\sum_{k=1}^N\rightarrow \int\limits_0^{2\pi}
\sigma_1(\theta)\, d\theta.$$
At the same time
$${\partial^2 {\tilde F}_N\over \partial \theta_k^{(1)2}}\sim
{\cal O}\Bigl({1\over N}\Bigr)$$
and can be neglected. As to the last term in \eqfn,
$$\eqalign{
&{1\over N^3} \sum_{k=1}^N {1\over {\cal D}(\theta^{(1)})}
{\partial^2\over
\partial \theta_k^{(1)2}} {\cal D}(\theta^{(1)})=
{1\over N^2}\sum_{k=1}^N{1\over {\cal D}(\theta^{(1)})}
{\partial\over \partial \theta^{(1)}_k}\Bigl[U_k
{\cal D}(\theta^{(1)})\Bigr]\cr&=
{1\over N^2}\sum_{k=1}^N\Bigl[{\partial U_k\over
\partial \theta^{(1)}}+U_k^2\Bigr]=
{1\over N^3}\sum_{j, k\atop j\ne k}{1\over 4 \sin^2 {\theta_k^{(1)}
-\theta_j^{(1)}\over 2}}+{1\over N}\sum_{k=1}^N U_k^2.\cr}$$
When $N\to \infty$
$$U_k \to U(\theta)={1\over 2}-\kern-1.12em\int
{\rm cotan}\biggl({\theta_k-\theta^{\prime}\over 2}\biggr)\,
\sigma_1(\theta^{\prime})\, d\theta^{\prime}$$
and
$${1\over N}\sum_{k=1}^N U_k^2 = \int\limits_0^{2\pi} U^2(\theta)\,
\sigma_1(\theta)\, d\theta.$$
As to the term $${1\over N^3} \sum_{j\ne k}
{1\over 4 \sin^2 {\theta_k^{(1)}-\theta_j^{(1)}\over 2}}$$
it might appear that it is ${\cal O}(1/N)$, since it contains only
two summations (each of which naively contributes a factor of $N$)
divided by $N^3$. However, due to the singularity at $\theta_k=
\theta_j$ this term is in fact not negligible. Its ${\cal O}(1)$
contribution comes from the regions where $|k-j|\ll N$ so that
$\theta_k^{(1)}-\theta_j^{(1)}\simeq (k-j)/(N\sigma_1(\theta_k^{(1)}))$
and
$$\eqalign{
&{1\over N^3} \sum_{j, k\atop j\ne k}
{1\over 4 \sin^2 {\theta_k^{(1)}-\theta_j^{(1)}\over 2}}=
{1\over N^3} \sum_{j, k\atop j\ne k}
{1\over (\theta_k^{(1)}-\theta_j^{(1)})^2}=
{1\over N^3}\sum_{k=1}^N\biggl[\sum_{j=1\atop j\ne k}^N
{N^2 \sigma_1^2\bigl(\theta_k^{(1)}\bigr)\over (j-k)^2}\biggr]
\cr&={1\over N}\sum_{k=1}^N{\pi^2\over 3}\sigma_1^2(\theta_k^{(1)})=
{\pi^2\over 3}\int\limits_0^{2\pi}\sigma_1^3(\theta) \, d\theta\cr}$$
where we used the identity
$$\sum_{j\ne k}{1\over (j-k)^2}={\pi^2\over 3}.$$

Combining the pieces we obtain
\eqn\flarrlim{
2{\partial {\tilde F}\over \partial A}=
\int\limits_{0}^{2\pi}\sigma_1(\theta)\, d\theta \biggl\{
\biggl({\partial\over \partial\theta}{\delta {\tilde F}\over
\delta\sigma_1(\theta)}\biggr)^2+ 2 U(\theta)
{\partial\over \partial\theta}{\delta {\tilde F}\over
\delta\sigma_1(\theta)}+U^2(\theta)+ {\pi^2\over 3}\sigma_1^2(\theta)
 \biggr\}.}
If we now introduce a new functional $S\bigl[\sigma_1(\theta),
\sigma_2(\theta)\bigl| A\bigr]$ according to \actrep\ then
$${\partial\over \partial\theta}{\delta {\tilde F}\over
\delta\sigma_1(\theta)}+U(\theta)=
{\partial\over \partial\theta}{\delta S\over
\delta\sigma_1(\theta)}$$
so that \flarrlim\ entails \hamjac.

Let us emphasize an important aspect of this derivation.
In making the transition from \eqfn\ to \flarrlim\ we relied on the
fact that the large $N$ limit ${\tilde F}= \lim_{N\to\infty}
{\tilde F}_N$ exists. The same would not be true if we had attempted
a shortcut, trying to replace ${\cal D}(\theta^{(1)})
{\tilde {\cal Z}}_N$ by $\exp(N^2 {\tilde G}_N)$ in \predas. In fact,
the functional ${\cal D}(\theta^{(1)})
{\tilde {\cal Z}}_N$ is not positive definite and therefore
${\tilde G}_N$ does not have a well defined large $N$ limit. The functional
$S$ introduced in \actrep\ is related to ${\tilde {\cal Z}}_N$
by $\exp(N^2S)=|{\cal D}(\theta^{(1)})||{\cal D}(\theta^{(2)})|
{\tilde {\cal Z}}_N$, the absolute value signs being essential. In fact,
it is the difference between ${\cal D}(\theta^{(1)})$ and
$|{\cal D}(\theta^{(1)})|$ that gives rise to the interaction term
${\pi^2\over 3}\int\sigma_1^3(\theta)\, d\theta$ in \hamjac.

\appendix{B}{The large $N$ limit of the Itzykson--Zuber integral.}

The large $N$ limit of the Itzykson--Zuber integral \itzyk\ can
be studied
using esentially the same technique that we used in appendix A to
obtain the collective field theory of QCD. First, one represents
$I_N$ in the form
\eqn\izzuber{
I_N(A, B)={{\rm det}|\!|\ee^{N a_k b_j}|\!|\over \Delta(a_k)\Delta(b_j)}
=\ee^{{N\over 2}\bigl[\sum_{k=1}^{N}a_k^2 +\sum_{k=1}^N b_j^2\bigr]}
{{\rm det}|\!|\ee^{-{N\over 2}(a_k- b_j)^2}|\!|
\over \Delta(a_k)\Delta(b_j)}.}
Then it is easy to check directly that the quantity
\eqn\jayn{J_N(t|A, B)={1\over t^{N^2\over 2}}
{{\rm det}|\!|\ee^{-{N\over 2t}(a_k- b_j)^2}|\!|
\over \Delta(a_k)\Delta(b_j)}}
satisfies the partial differential equation
\eqn\pardejayn{
2N{\partial J_N\over \partial t}=
{1\over \Delta(a)}\sum_{i=1}^N {\partial^2\over \partial a_i^2}
\bigl[\Delta(a)\,  J_N\bigr].}
Once $J_N$ is known, $I_N$ can be retrieved as
$$I_N=J_N(t=1)\, \exp\biggl[{{N\over 2}\Bigl(
\sum_{k=1}^{N}a_k^2 +\sum_{k=1}^N b_j^2\Bigr)}\biggr].$$
Equation \pardejayn\ which is analogous to \predas\ can be treated by methods
described in appendix A. The only difference is that the quantity
$U_k$ is given not by \cottang\ but by
\eqn\invlin{
U_k\equiv {1\over N} {\partial\over \partial a_k}\ln \Delta(a)
={1\over N}\sum_{j\ne k}{1\over a_k-a_j}}
with the large $N$ limit
$$U(a)=-\kern-1.12em\int {\alpha(a^{\prime})\, da^{\prime}\over
a- a^{\prime}}$$
where $\alpha(a)$ is the density of the eigenvalues $\{a_k\}$.
Performing the transformations described in appendix A one
deduces that the functional
\eqn\functio{\eqalign{
S[t|\alpha, \beta]=\lim_{N\to\infty}\biggl\{
{1\over N^2} \ln J_N(t|A, B)\biggr\}
+&{1\over 2}\int \alpha(a)\,\alpha(a^{\prime})
\, \ln|a-a^{\prime}|\, da\, da^{\prime}\cr&+
{1\over 2}\int \beta(b) \, \beta(b^{\prime})\,
\ln|b- b^{\prime}|\, db\, db^{\prime}\cr}}
satisfies the following differential equation
\eqn\fdreq{
{\partial S\over \partial t}=
{1\over 2}
\int\limits_{-\infty}^{+\infty} \alpha(a) \Biggl[
\biggl({\partial\over \partial a}
{\delta S\over \delta \alpha(a)}\biggr)^2-{\pi^2\over 3}
\alpha^2(a)\Biggr]\, da.
}

Like \hamjac\ this is the Hamilton--Jacobi equation for
the dynamical system with the Hamiltonian
\eqn\hmm{
H\bigl[\rho(a), \Pi(a)\bigr]={1\over 2}
\int\limits_{-\infty}^{+\infty}\rho(a)\biggl[\biggl({\partial \Pi
\over \partial a}\biggr)^2- {\pi^2\over 3}\rho^2(a)\biggr]\, da
}
However, as opposed to \hamjac, in this dynamical system
$\rho(a)$ is the distribution on an infinite real line
rather than on a circle. In fact, it is the noncompactness
of support of eigenvalue densities that distinguishes
the Itzykson--Zuber integral from the large $N$ QCD.

Equation \fdreq\ is very hard, if not impossible, to
solve in general. Fortunately, however, it is easy to
find its particular solution which describes
the large $N$ limit of the Itzykson--Zuber integral \REFmat.
This solution is given by the action of dynamical system
\hmm\ along its classical trajectory connecting the densities
$\rho(a)=\alpha(a)$ and $\rho(b)=\beta(b)$ within time $t$.
It is not difficult to show that such quantity
(which is known in classical mechanics as the principal
Hamilton function) does indeed satisfy \fdreq. Moreover, the
variational derivatives at the end of the trajectory
equal the corresponding canonical momenta:
$${\delta S\over \delta \alpha(a)}=\Pi(a, t=0), \qquad
{\delta S\over \delta \beta(b)}=-\Pi(b, t=1).$$
To find the above mentioned trajectory we must solve the equations
of motion which follow from the Hamiltonian \hmm. As in \euler,
\hopf\ these can be conveniently transformed into a single equation
for the quantity $f(t, a)={\partial \Pi(a)\over \partial a}
+i\pi \rho(a)$:
\eqn\hopfa{
{\partial f\over \partial t}+f{\partial f\over \partial a}=0.}
Then
$${\partial\over \partial a}{\delta S\over \delta \alpha(a)}=
{\partial \Pi(a, t=0)\over \partial a}={\rm Re}\, f(t=1, a).$$

Now we are able to derive the constraints \pairf.
We notice that the general solution of \hopfa\ can be written down
in the parametric form
\eqn\param{
\left\{\eqalign{
&x=R(\xi)+F(\xi) \, t\cr
&f(x, t)= F(\xi)\cr}\right.}
where $R(\xi)$ and $F(\xi)$ are some functions of the formal parameter
$\xi$. These functions should be determined from the initial conditions.
The conditions to be imposed in our case are
${\rm Im}\, f(t=1, a)=\pi\alpha(a)$ and ${\rm Im}\, f(t=0, b)=\pi\beta(b)$.
If we introduce the two analytic functions
$G_+(x)$ and $G_-(x)$ according to
\eqn\accord{
\left\{\eqalign{
&G_+(x)=x+f(t=0, x)\cr
&G_-(x)=x-f(t=1, x)\cr}\right.}
then ${\rm Im}\, G_+(x)=\pi\alpha(x)$ and ${\rm Im}\, G_-(x)=
-\pi\beta(x)$. In addition, from \param\ we deduce
$$G_+(x)=x+[ F\circ R^{-1}](x)=
\bigl[(F+R)\circ R^{-1}\bigr](x)$$
as well as
$$G_-(x)=x- \bigl[F\circ(F+R)^{-1}\bigr](x)=
\bigl[R\circ(F+R)^{-1}\bigr](x)$$
where $\circ$ denotes functional composition. Therefore,
$$G_+\bigl(G_-(x)\bigr)=G_-\bigl(G_+(x)\bigr)=x$$
completing the proof of \pairf.

\appendix{C}{Wilson loops on a plane and the Hopf equation.}

In this appendix we will review the derivation of Wilson loops on
a plane which uses the method of loop equations \REFkazkos\REFrossi.
Using this example
we will be able to see the close connection existing between
the Hopf equation \hopf\ and the Kazakov--Kostov equations for
the two-dimensional QCD. In addition we will derive the explicit
expressions \wlplane\ for Wilson loops on a plane.

\ifig\reconec{The reconnection of a Wison loop at the self-intersection
point employed in the two-dimensional loop equations.}
{\epsfxsize2.25in\epsfbox{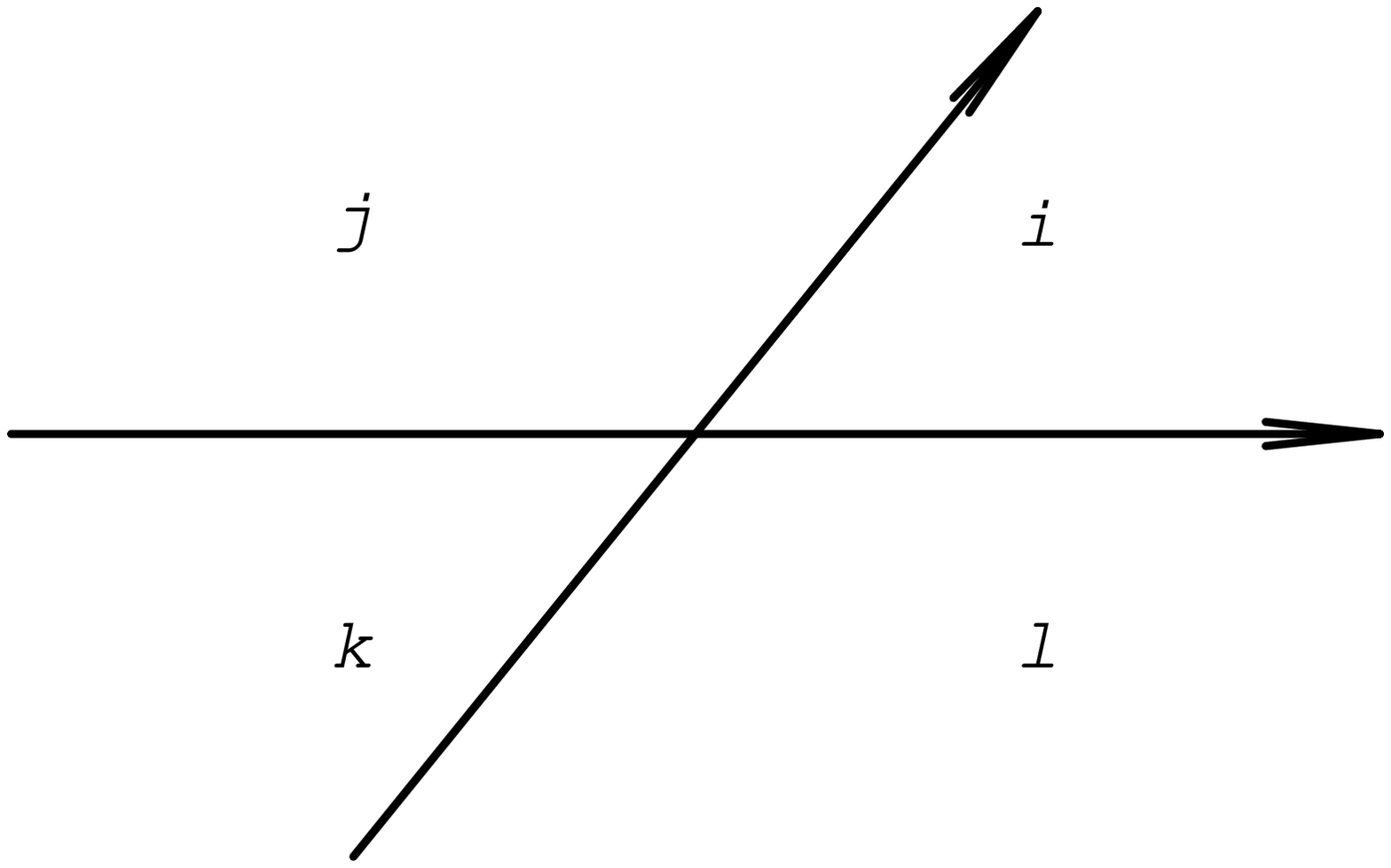}\hskip0.1in
\epsfxsize2.25in\epsfbox{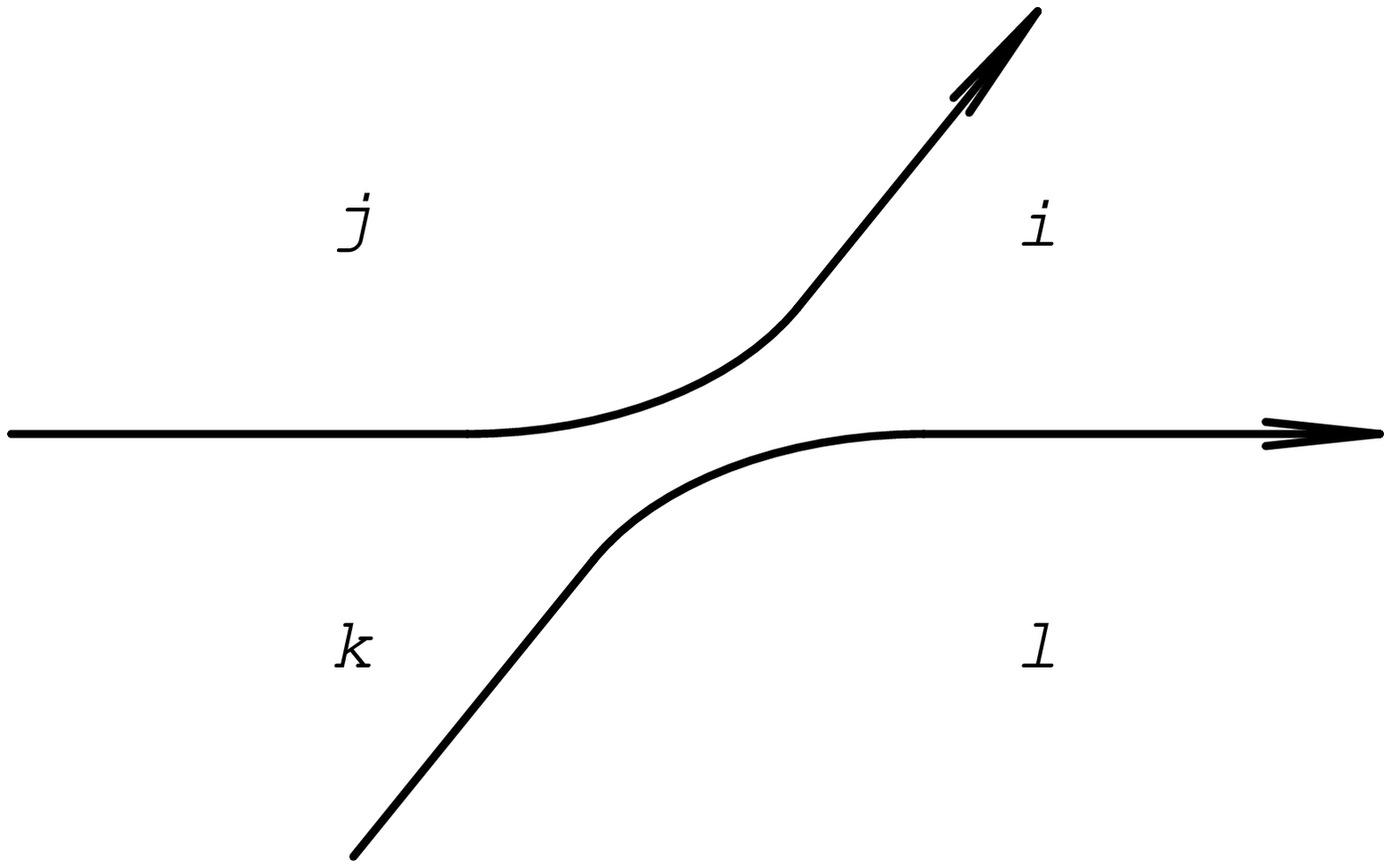}}

Generally, the loop equations relate the value of a self-intersecting
Wilson loop $W$ (fig.7) to the values of Wilson loops with
smaller number of self-intersections. In particular, if we
``reconnect" the self-intersection on the left in  \reconec\ to obtain the
contour shown on the right  then the Wilson loop $W$ for the
original contour satisfies the Kazakov--Kostov equation \REFkazkos
\eqn\kazkos{
{\hat R}W= W^{\prime}W^{\prime\prime}}
where the operator ${\hat R}$ is given by\foot{We remember that
in the two-dimensional QCD the Wilson loops depend only on the
areas bounded by the contour but not on the contour shape. This
follows from the invariance of ${\rm QCD}_2$ with respect to the
group of area preserving diffeomorphisms.}
\eqn\operr{
{\hat R}={\partial\over \partial S_k}+
{\partial\over \partial S_i}-{\partial\over \partial S_l}
-{\partial\over \partial S_j}.}
In  this formula $S_{\alpha}$ is the area of the window marked
by the subscript $\alpha\in\{i, j, k, l\}$.

\ifig\selfis{A Wilson loop with two self-intersections (this corresponds
to the winding number equal to 3). The areas of domains bounded by the loop
are denoted as $S_0, S_1, S_2$, counting from the outside.}
{\epsfxsize3.0in\epsfbox{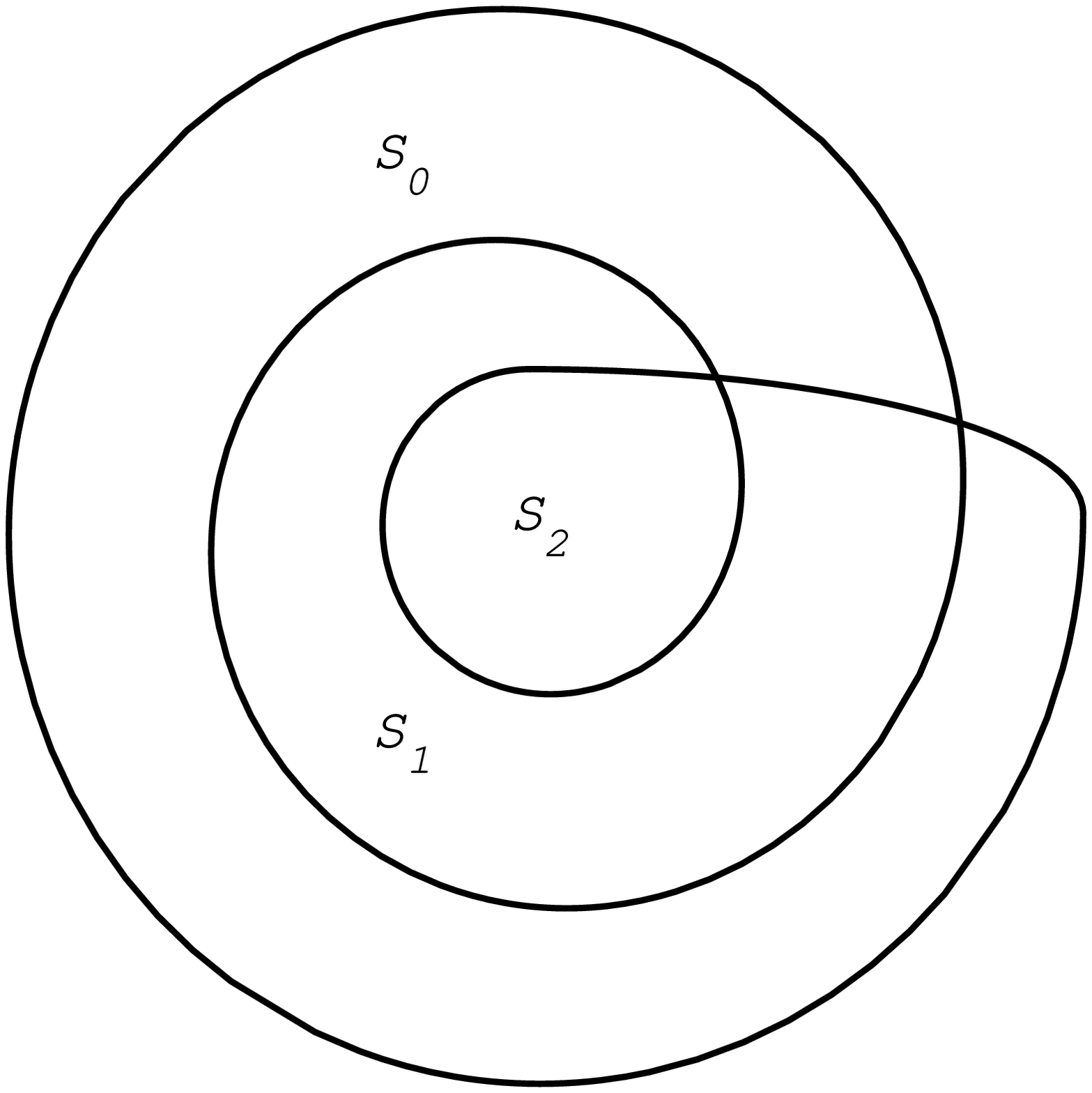}}

Below we will be interested in the loops which wind several times
around the circle of area $A$. In \selfis\ this would correspond to a
situation when $S_0, S_1, \dots, S_{n-1}\to 0, \ S_n\to A$.
In this case the Wilson loop averages $W_n(A)$ depend only on
one variable $A$ and, due to \kazkos, satisfy the equations
\eqn\lpe{
\eqalign{
&\Bigl(2{\partial\over \partial S_0}-
{\partial\over \partial S_1}\Bigr)W_n= W_{n-1} W_0\cr
&\Bigl(2{\partial\over \partial S_1}-
{\partial\over \partial S_0}-{\partial\over \partial S_2}\Bigr)W_n= W_{n-2}
W_1\cr
&\quad\vdots\cr
&\Bigl(2{\partial\over \partial S_k}-
{\partial\over \partial S_{k-1}}-{\partial\over \partial S_{k+1}}\Bigr)W_n=
W_{n-k-1}W_{k}\cr
&\quad\vdots\cr
}}
These equations can be simplified if we introduce a new set
of variables $A_n=S_n, \ A_{n-1}=A_n+S_{n-1},
\dots, A_k=A_{k+1}+s_k, \dots$. Then $\partial/\partial A_0=
\partial/\partial S_0$ and $\partial/\partial A_k=
\partial/\partial S_k-\partial/\partial S_{k+1}$ so that from
\lpe\ we obtain
$$\Bigl({\partial\over \partial A_k}-
{\partial\over \partial A_{k+1}}\Bigl)W_n=
W_{n-k-1}W_k,$$
and, consequently,
\eqn\lpa{
\Bigl({\partial\over \partial A_0}-
{\partial\over \partial A_{k}}\Bigr)
W_n(A_0, A_1, \dots, A_n)=
\sum_{l=0}^{k-1}W_{n-k-1}(A)W_k(A).}
Now, if $S_0, S_1, \dots, S_{n-1}\to 0$ then
$A_0=A_1=\dots=A_n=A$ and
\eqn\difa{{d W_n(A)\over dA}=
\sum_{k=0}^n\biggl[{\partial\over \partial A_{k}}
W_n(A_0, A_1, \dots, A_n)\biggr]\biggr|_{{\rm all} A_k=A}.}

Let us also recall that the dependence of $W_n$ on  the external area $A_0$
(or, the same, $S_0$) is always exponential, $W_n\sim \ee^{-A_0/2}$
so that
$${\partial W_n\over \partial A_0}=-{1\over 2}W_n.$$
This, together with \lpa\ and \difa\ allows us to write down a
recursion relation for $W_n$:
\eqn\rrel{
-{dW_n\over dA}= {n+1\over 2}W_n+\sum_{l=0}^{n-1}(n-l)
W_lW_{n-l-1}.}
Its solution has the general form
$$W_n(A)=P_n(A)\,  \ee^{-{n+1\over 2}A}$$
where $P_n(A)$ is a polynomial in $A$ of degree $n$. It is easy to
show by induction that these polynomials are given by \laguerrep\
yielding the formula \wlplane\ \REFrossi.

Our goal here is different. We would like to see how this formalism
corresponds with the collective field description of QCD and, particularly,
with the Hopf equation \hopf.

In fact, the recursion relation \rrel\ resembles the Fourier transform
of the Hopf equation. Indeed, if we set
$$\psi(A, t)\equiv \sum_{n=0}^{\infty}W_n(A)\, \ee^{(n+1)({A\over 2}
+\zeta)}$$
then \rrel\ entails
$${\partial\psi\over \partial A}+\psi{\partial\psi\over \partial \zeta}
=0.$$
However, $\psi$ is not the same as the collective field function $f$
because by construction $\psi$ is real while $f$  has an imaginary part,
${\rm Im}\, f(t, \theta)=\pi\sigma_*(t, \theta)$.
But if we put $\zeta=-i\theta-{A\over 2}$ then
$${\rm Re}\, \psi\Bigl(A, -i\theta- {A\over 2}\Bigr)=
\sum_{n=0}^{\infty} W_n(A)\,\cos(n+1)\theta=
\pi\sigma_*(t, \theta)-{1\over 2}$$
where we have used the relation between $W_n$ and $\sigma_*$
$$W_n(C)=\int \limits_0^{2\pi} \sigma_*(t=A_1,\theta)\,
\ee^{i n \theta}\, d\theta$$
(see \wla).

Thus the function
\eqn\fhp{
f(A, \theta)=i\biggl[\psi\Bigl(A, -i\theta-{A\over 2}\Bigr)
+{1\over 2}\biggr]}
has the correct imaginary part, ${\rm Im}\, f(A, \theta)=
\pi\sigma_*(A, \theta)$. Moreover, since
$$\Bigl({\partial\over \partial\zeta}\Bigr)_A=i
\Bigl({\partial\over \partial\theta}\Bigr)_A, \quad
\Bigl({\partial\over \partial A}\Bigr)_{\zeta}=
\Bigl({\partial\over \partial A}\Bigr)_{\theta}+{i\over 2}
\Bigl({\partial\over \partial\theta}\Bigr)_A$$
the Hopf equation for $\psi$ transforms into the Hopf equation for
$f$:
\eqn\fhf{
\Bigl({\partial f\over \partial A}\Bigr)_{\theta}+
f\Bigl({\partial f\over \partial\theta}\Bigr)_A=0.}
Unlike $\psi$ the function $f$, which has both real and imagnary parts,
is precisely the function used in the collective field theory. Thus,
the formula
\fhp\ establishes the agreement between  the Hopf equation and the
loop equations of two-dimensional QCD.

\listrefs
\end

$$\eqalign{
2 {\partial {\tilde F}\over \partial A}=&\int\limits_0^{2\pi}\sigma_1(\theta)
\Biggl[\biggl({\partial\over \partial \theta}{\delta {\tilde F}\over
\delta \sigma_1 (\theta)}\biggr)^2+
2 \biggl({\partial\over \partial \theta}{\delta {\tilde F}\over
\delta \sigma_1 (\theta)}\biggr)\biggl({1\over 2}{\bf -}\kern-1.1em\int
\sigma_1 (\theta)\thinspace
{\rm cot}{\theta - \varphi\over 2}\thinspace d \varphi\biggr)\cr
+&{1\over 4}\biggl({\bf -}\kern-1.1em\int \sigma_1 (\theta)\thinspace
{\rm cot}{\theta - \varphi\over 2}\thinspace
d \varphi\biggr)^2\Biggr]\, d\theta
-{\pi^2\over 3}\int\limits_0^{2\pi}\sigma_1^3(\theta)\, d\theta.\cr
}$$
We have taken into account the remarkable identity
$$\lim\limits_{N\to\infty}{1\over N^3}\sum_{j\ne k}{1\over 4\sin^2
{\theta_j-\theta_k\over 2}}={\pi^2\over 3}\int\limits_0^{2\pi}
\sigma_1^3(\theta)\, d\theta$$
which holds in the large $N$ limit  and replaced
$$N{\partial {\tilde F}_N\over \partial \theta_j^{(1)}}\rightarrow
{\partial\over \partial\theta}{\delta {\tilde F}\over \delta \sigma_1(\theta)}
\biggr|_{\theta=\theta_j^{(1)}}\; ,\qquad
N{\partial^2 {\tilde F}_N\over \partial \theta_j^{(1)2}}\rightarrow
{\cal O}\Bigl({1\over N}\Bigr)$$
as $N\to\infty$. Finally, it follows that
\eqn\actrep{\eqalign{
&{\tilde F}\bigl[\sigma_1(\theta), \sigma_2(\theta)\bigl| A\bigr]=
S\bigl[\sigma_1(\theta), \sigma_2(\theta)\bigl| A\bigr]\cr
&-{1\over 2}\int \sigma_1(\theta)\sigma_1(\varphi)\ln\left|\sin
{\theta-\varphi\over 2}\right|\, d\theta\, d\varphi
-{1\over 2}\int \sigma_2(\theta)\sigma_2(\varphi)\ln\left|\sin
{\theta-\varphi\over 2}\right|\, d\theta\, d\varphi
\cr}}
where the functional $S$ is a solution of the Hamilton--Jacobi
equation
\eqn\hamjac{
{\partial S\over \partial A}={1\over 2}\int\limits_0^{2\pi}
\sigma_1(\theta)\Biggl[\biggl({\partial\over \partial\theta}
{\delta S\over \delta \sigma_1(\theta)}\biggr)^2-{\pi^2\over 3}
\sigma_1^2(\theta)\Biggr]
}
with $A$ playing the role of time, $\sigma_1(\theta)$ --the canonical
coordinate and $\Pi(\theta)\equiv\delta S/\delta \sigma_1(\theta)$ --the
momentum.

In fact, the required solution is easy to construct. To do this, we
solve the Hamilton equations of motion for the Hamiltonian
\eqn\hamm{
H\bigl[\sigma(\theta), \Pi(\theta)\bigr]={1\over 2}
\int\limits_0^{2\pi}\sigma(\theta)\biggl[\biggl({\partial \Pi
\over \partial \theta}\biggr)^2- {\pi^2\over 3}\sigma^2(\theta)\biggr]
}
and pick up the particular solution which satisfies the boundary conditions
\eqn\bc{
\left\{\eqalign{\sigma(\theta)\bigr|&_{t=0}=\sigma_1(\theta)\cr
\sigma(\theta)\bigr|&_{t=A}=\sigma_2(\theta)\cr}\right.
}
where, as before, $\sigma_1$ and $\sigma_2$ are the eigenvalue densities of
matrices $U_{C_1}$ and $U_{C_2}$. Then $S[\sigma_1, \sigma_2|A]$
equals the classical action calculated for this particular
solution\foot{If there are several such solutions, the one with
the largest value of $S$ must be chosen.}.

The equations of motion which follow from \hamm\ are
\eqn\euler{
\left\{\eqalign{&{\partial \sigma\over \partial t}+
{\partial \over \partial \theta}(\sigma v)=0 \cr
&{\partial v\over \partial t}+v{\partial v \over \partial \theta}=
 {\partial \over \partial \theta}\left({\pi^2\over 2}\sigma^2\right)
\cr}\right. \, , \quad v(\theta)\equiv {\partial \Pi\over \partial\theta}.
}
These are the Euler equations for the fluid with pressure
$P=-{\pi^2\over 2}\sigma^2$. The solution we are looking for
corresponds to the process where the density profile $\sigma_1(\theta)$
evolves into the profile $\sigma_2(\theta)$ during the time equal to $A$.

Unlike the free energy ${\tilde F}$, the action $S$ has the important
property of additivity.

\end